\documentclass[a4paper,12pt]{article}
\usepackage{amsmath,amssymb,mathtools}
\usepackage{jheppub}
\pdfoutput=1
\usepackage{graphicx}
\usepackage{hyperref}
\usepackage[utf8x]{inputenc}
\usepackage[normalem]{ulem}
\usepackage{slashed}
\usepackage{xcolor,cancel}
\usepackage{array}
\newcommand{\PreserveBackslash}[1]{\let\temp=\\#1\let\\=\temp}
\newcolumntype{C}[1]{>{\PreserveBackslash\centering}p{#1}}

\newcommand{\ord}[1]{\mathcal{O}\left({#1}\right)}
\newcommand{\diff}{\mathrm{d}}

\title{Assessment of the Dimension-5 Seesaw Portal and Impact of Exotic Higgs Decays on Non-Pointing Photon Searches}

\author[a]{F.~Delgado}
\author[b,1]{L.~Duarte\note{Corresponding authors.}}
\author[a,1]{J.~Jones-P\'erez}
\author[a]{C.~Manrique-Chavil}
\author[a]{S.~Peña}

\affiliation[a]{Secci\'on F\'isica, Departamento de Ciencias, Pontificia Universidad Cat\'olica del Per\'u, Apartado 1761, Lima, Peru}
\affiliation[b]{Instituto de F\'isica, Facultad de Ciencias, Universidad de la República, Iguá 4225, Montevideo, Uruguay}

\emailAdd{a20130201@pucp.edu.pe}
\emailAdd{lucia@fisica.edu.uy}
\emailAdd{jones.j@pucp.edu.pe}
\emailAdd{cristian.manrique@pucp.edu.pe}
\emailAdd{a20120270@pucp.edu.pe}

\abstract{
The Dimension-5 Seesaw Portal is a Type-I Seesaw model extended by $d=5$ operators involving the sterile neutrino states, leading to new interactions between all neutrinos and the Standard Model neutral bosons. In this work we focus primarily on the implications of these new operators at the GeV-scale. In particular, we recalculate the heavy neutrino full decay width, up to three-body decays. We also review bounds on the dipole operator, and revisit LEP constraints on its coefficient. Finally, we turn to heavy neutrino pair production from Higgs decays, where the former are long-lived and disintegrate into a photon and a light neutrino. We probe this process by recasting two ATLAS searches for non-pointing photons, showing the expected event distribution in terms of arrival time $t_\gamma$ and pointing variable $|\Delta z_\gamma|$.
}

\begin{document}

\maketitle
\flushbottom

\section{Introduction}
\label{sec:notrev}

The simplest Standard Model (SM) extensions that can account for neutrino masses, and hence the neutrino oscillations phenomena, require the introduction of sterile right-handed neutrinos $\nu_R$. These SM singlets allow Dirac masses for the light neutrinos we know. However, more appealing mechanisms for light neutrino mass generation are available if the $\nu_R$ have Majorana masses. Typical examples are the Type I Seesaw \cite{Minkowski:1977sc, Mohapatra:1979ia, Yanagida:1980xy, GellMann:1980vs, Schechter:1980gr} as well as the linear and inverse Seesaw mechanisms~\cite{Malinsky:2005bi,Mohapatra:1986bd}. 

Even though the $\nu_R$ fields are gauge singlets, they can have interactions with other SM particles. The first possibility comes after diagonalizing the Seesaw $\nu_L-\nu_R$ mass matrix, such that the heavy states $N_h$, which are mostly $\nu_R$, can mix with the active $\nu_L$. The $N_h$ are then no longer sterile, behaving as very weakly coupled particles, with consequences in collider searches and low energy processes like neutrinoless double beta decay~\cite{Atre:2009rg,Deppisch:2015qwa,Abdullahi:2022jlv}. A second possibility that does not necessarily involve the Seesaw arises when including the $\nu_R$ fields in an effective field theory (EFT) as low energy degrees of freedom. This is the Standard Model Effective Field Theory framework extended with right-handed neutrinos, $\nu_R$SMEFT\footnote{Also called SMNEFT and $N_R$SMEFT  in the literature.}, with operators known up to dimension $d=9$~\cite{delAguila:2008ir, Aparici:2009fh, Liao:2016qyd, Bhattacharya:2015vja, Li:2021tsq}. Here, one can proceed as in~\cite{Terol-Calvo:2019vck, Escrihuela:2021mud} and use the so-called neutrino non-standard interactions (NSI) and general neutrino interactions (GNI) to constrain the $\nu_R$SMEFT~\cite{Bischer:2019ttk, Han:2020pff}. Bounds can also be placed from beta decays~\cite{Falkowski:2020pma}.

In this sense, combinations of the Type-I Seesaw and higher dimension effective interactions for the $\nu_R$ have reached far attention, since the $N_h$ production rates and decay widths can be drastically changed by the effective interactions. This can lead to a variety of signals that can be studied at different experiments. For example, for dimension-6 operators, the $\nu_R$SMEFT lagrangian includes novel four-fermion operators as well as $N_h$ interactions with the Higgs and the standard vector bosons. The phenomenology of these $d=6$ heavy $N_h$ interactions, when dominant with respect to those involving the mixing with the active neutrino states, has been studied extensively in processes at the LHC and future far detectors in~\cite{delAguila:2008ir, Duarte:2016caz, Alcaide:2019pnf, Butterworth:2019iff, Beltran:2021hpq, Cottin:2021lzz}. Constraints and prospects on the sensitivity to the effective Wilson couplings from rare meson and tau decays into heavy neutrinos have also been studied in~\cite{Duarte:2019rzs, Duarte:2020vgj,  DeVries:2020jbs, Biekotter:2020tbd, Zhou:2021ylt}, as well as their contributions to leptonic anomalous magnetic moments \cite{Cirigliano:2021peb}. Prospects for the production and sensitivity to $d=6$ interactions in future $e^+e^-$ and $e^-p$ colliders have been considered in \cite{Duarte:2018kiv, Duarte:2018xst,Barducci:2022hll, Zapata:2022qwo}.

In front of this, we want to focus on the phenomenology of the Type-I Seesaw, extended by dimension-5 operators, which has somewhat received less attention. Our final objective is to analyze scenarios where the heavy neutrinos can be long-lived, decaying into a photon and light neutrino, and on the possibility of detecting them at the LHC. This means that the $N_h$ will have masses around and above the GeV scale.

In our setup, we extend the SM by adding three sterile neutrinos $\nu_R$. The renormalizable part of the lagrangian is:
\begin{equation}
 \mathcal L=\mathcal L_{SM}+i\bar\nu_{Rs}\,\slashed\partial\,\nu_{Rs}-\left(\bar L_a (Y_\nu)_{a s}\,\tilde\phi\, \nu_{Rs}
 +\frac{1}{2}\bar\nu_{Rs} (M_N)_{ss'}\nu^c_{Rs'}+h.c.\right)
\end{equation}
where $a=e,\,\mu,\,\tau$ and $s,s'=s_1,s_2,s_3$, and the matrix $M_N$ is symmetric. When allowing for $d=5$ operators involving the new neutrino states, one finds the following terms:
 \begin{equation}
 \label{eq:LagEff}
  \mathcal L_5=\frac{(\alpha^\dagger_W)_{ab}}{\Lambda}(\bar L_a\tilde\phi)(\phi^\dagger \tilde L^c_b)
  +\frac{(\alpha_{N\phi})_{ss'}}{\Lambda}(\phi^\dagger\phi)\,\bar\nu_{Rs}\,\nu^c_{Rs'}
  +\frac{(\alpha_{NB})_{ss'}}{\Lambda}\bar\nu_{Rs}\,\sigma^{\mu\nu}\nu^c_{Rs'}\,B_{\mu\nu}+h.c.
 \end{equation}
where $\tilde L^c=\varepsilon L^c$ and $\sigma^{\mu\nu}=\tfrac{i}{2}[\gamma^\mu,\,\gamma^\nu]$. The first term corresponds to the well-known Weinberg operator~\cite{Weinberg:1979sa}, the second operator was introduced by Anisimov~\cite{Anisimov:2006hv} and Graesser~\cite{Graesser:2007yj,Graesser:2007pc},  while the third one, which is a dipole operator, was studied by Aparici, Kim, Santamaria and Wudka~\cite{Aparici:2009fh}. Again, $\alpha_W$ and $\alpha_{N\phi}$ must be symmetric matrices, while $\alpha_{NB}$ must be antisymmetric. The Anisimov-Graesser and dipole operators allow for interactions of the heavy neutrinos with the Higgs, photon and $Z$ which are not suppressed by the active-heavy mixing. 

Since we are ultimately interested in $N_h$ decays involving photons, the dipole operator will play a central role in our research. It is worth noting that in order to have a non-vanishing dipole coefficient, one requires at least two $\nu_R$ states. Thus, simplified approaches considering only one sterile state forbid the presence of this antisymmetric operator, pushing the dipole interaction to $d=6$ terms. Such studies have been done in~\cite{Magill:2018jla,Brdar:2020quo}, which consider a heavy Dirac neutrino with dipole interactions to a SM neutrino and the photon. The papers obtain constraints on this dipole operator from the intensity, energy and cosmic frontiers, as well as presenting future prospects. Although presented in a $d=6$ electroweak invariant realization, these bounds can also be interpreted in a $d=5$ framework.

Other works considering at least two GeV-scale sterile states and Seesaw mixings mostly focus on the study of the Anisimov-Graesser operator and its Higgs phenomenology at the LHC~\cite{Aparici:2009fh,Graesser:2007yj,Graesser:2007pc,Caputo:2017pit,Jones-Perez:2019plk}. In addition, apart from the original work of~\cite{Aparici:2009fh}, collider effects of the dipole operator have been considered at Higgs factories in~\cite{Barducci:2020icf}. Thus, to the best of our knowledge, the $d=5$ dipole operator has not received enough attention. In this sense, this work aims to help fill this gap, covering search strategies involving photons coming from heavy neutrino decay, concentrating on the novel signal of non-pointing photons from exotic Higgs decays at the LHC.

This paper is organized as follows. In Section~\ref{sec:massparam} we define our notation and conventions for all neutrino masses and mixing. Section~\ref{sec:Couplings} translates the operators in Eq.~(\ref{eq:LagEff}) into couplings between the mass eigenstates. We also carry out a detailed study of how the partial widths, branching ratios and lifetimes of the heavy neutrinos are affected by the dipole operator. On Section~\ref{sec:Constraints} we review constraints on the latter, concentrating on those around the GeV scale. To this end, we revisit the LEP constraints on light-heavy neutrino production. Finally, in Section~\ref{sec:NonPointing} we turn towards the prospects of identifying long-lived GeV-scale heavy neutrinos decaying into photons, recasting two ATLAS non-pointing photon searches.

\section{Masses and Parametrization}
\label{sec:massparam}

As is well known, in the absence of effective operators, one can organize the active and sterile neutrino fields in a left-handed multiplet $\Psi_L\equiv(\nu_L,\,\nu_R^c)^T$, such that the neutrino mass matrix is written:
\begin{equation}
 \label{eq:numassmatrix}
 \mathcal{M}_\nu=\left(\begin{array}{cc}
0 & m_D \\
m_D^T & M_N
\end{array}\right)~,
\end{equation}
where $m_D\equiv\tfrac{v}{\sqrt2}Y^*_\nu$. The mass matrix is diagonalized by a unitary transformation:
\begin{equation}
 U^T\mathcal{M}_\nu\,U=\mathcal M_\nu^{\rm diag}~.
\end{equation}
The mixing matrix $U$ relates the interaction eigenstates $\Psi_{L\alpha}$ ($\alpha=e,\,\mu,\,\tau,\,s_1,\,s_2,\,s_3$), to the mass eigenstates $n_i$ ($i=1,\ldots,6$), that is, $\Psi_{L\alpha}=U_{\alpha i}\,n_{Li}$. In the following, our notation might distinguish between the lightest and heaviest neutrino mass eigenstates:
\begin{align}
 \nu_{\ell}=n_i & & i=\ell=1,\ldots,3 \\
 N_h=n_i & & i=h=4,\ldots,6
\end{align}
We refer to this scenario as the standard Seesaw.

When including effective operators, we find further contributions to the neutrino mass terms after electroweak symmetry breaking. If we define:
 \begin{align}
  \label{eq:defmLLmRR}
  m_{LL}=\frac{v^2\,\alpha_W}{\Lambda} & & m_{RR}=\frac{v^2\,\alpha_{N\phi}}{\Lambda}+M_N
 \end{align}
the neutrino mass matrix is now:
\begin{equation}
\label{eq:numassmatrixeff}
 \mathcal{M}_\nu=\left(\begin{array}{cc}
m_{LL} & m_D \\
m_D^T & m_{RR}
\end{array}\right)
\end{equation} 

In order to understand the structure of the diagonalization matrix $U$, we will first proceed as in~\cite{Ibarra:2010xw}. The idea is to do an initial rotation that block-diagonalizes the mass matrix:
\begin{equation}
\label{eq:ThetaDiag}
\begin{pmatrix}
I-\Theta^*\Theta^T/2 & -\Theta^* \\
\Theta^T & I-\Theta^T\Theta^*/2
\end{pmatrix}
\left(\begin{array}{cc}
m_{LL} & m_D \\
m_D^T & m_{RR}
\end{array}\right)
\begin{pmatrix}
I-\Theta\Theta^\dagger/2 & \Theta \\
-\Theta^\dagger & I-\Theta^\dagger\Theta/2
\end{pmatrix}
= 
\begin{pmatrix}
 M_{\rm light} & 0 \\
 0 & M_{\rm heavy}
\end{pmatrix}
\end{equation}
The $\Theta$ matrix is assumed to have small entries, such that at leading order in $\Theta$, block-diagonalization requires:
\begin{equation}
 m_D\approx\Theta^* m_{RR}-m_{LL}\,\Theta
\end{equation}
As $m_{LL}$ contributes directly to the light neutrino masses, one would expect it to be very small in front of $m_{RR}$. This means we can approximate $m_D\approx\Theta^* m_{RR}$, and thus:
\begin{eqnarray}
\label{eq:Mlight}
M_{\rm light}&=& -m_D \,m_{RR}^{-1}\,m_D^T+m_{LL} \\
\label{eq:Mheavy}
M_{\rm heavy}&=&m_{RR}+\frac{1}{2}\left(m_{RR}^{-1*}\,m_D^\dagger\, m_D+m_D^T\, m_D^*\, m_{RR}^{-1}\right)
\end{eqnarray}
The next step is to diagonalize the block-diagonal matrix:
\begin{equation}
\begin{pmatrix}
U_{\rm PMNS}^T & 0 \\
0 & V^T
\end{pmatrix}
\begin{pmatrix}
 M_{\rm light} & 0 \\
 0 & M_{\rm heavy}
\end{pmatrix}
\begin{pmatrix}
U_{\rm PMNS} & 0 \\
0 & V
\end{pmatrix}=
\begin{pmatrix}
\hat m_\ell & 0 \\
0 & \hat M_h
\end{pmatrix}~,
\end{equation}
where $\hat m_\ell={\rm diag}(m_1,\,m_2,\,m_3)$ and $\hat M_h={\rm diag}(M_4,\,M_5,\,M_6)$. If we decompose $U$ into four blocks:
\begin{equation}
 U=\left(\begin{array}{cc}
U_{a\ell} & U_{ah} \\
U_{s\ell} & U_{sh}
\end{array}\right)~,
\end{equation}
we can identify:
\begin{align}
 U_{a\ell} &= \left(I-\Theta\Theta^\dagger/2\right)U_{\rm PMNS},  & 
 U_{ah} &= \Theta\,V \nonumber \\
 U_{s\ell} &= -\Theta^\dagger U_{\rm PMNS}~, &
 U_{sh} &= \left(I-\Theta^\dagger\Theta/2\right)V~.
\end{align}
Thus, the matrices $\Theta^\dagger\Theta$ and $\Theta\Theta^\dagger$ are connected to the non-unitarity of the active-light and sterile-heavy mixing. Morover, the $V$ matrix is related to re-definitions of the sterile neutrinos, and in the absence of a preferred basis can be taken equal to the identity.

We will now focus on eliminating the $\Theta$ matrix in favour of a more useful parametrization, such as~\cite{Casas:2001sr,Donini:2012tt}, which guarantees that the measured $\nu_\ell$ masses are always reproduced. To do so, we rewrite $M_{\rm light}$ and keep terms of leading order in $\Theta$ (or equivalently, in $m_D$):
\begin{eqnarray}
 U_{\rm PMNS}^* \,\hat m_\ell\, U^\dagger_{\rm PMNS}&=&-m_D \,m_{RR}^{-1}\, m_D^T+m_{LL} \nonumber \\
 &=&-(\Theta V)^*\hat M_{h}(\Theta V)^\dagger+m_{LL}
 \end{eqnarray}
 If we now define $\Delta_\ell\equiv \left(I-\hat m_\ell^{-1/2}\,U^T_{\rm PMNS}\,m_{LL}\,U_{\rm PMNS}\,\hat m_\ell^{-1/2}\right)$, we can write the equation above as:
\begin{equation}
U_{\rm PMNS}^* \,\hat m_\ell^{1/2}\,\Delta_\ell\,\hat m_\ell^{1/2}\, U^\dagger_{\rm PMNS}=-(\Theta V)^*\hat M_{h}(\Theta V)^\dagger
\end{equation} 
which in turn allows us to define the complex matrix $R$:
\begin{equation}\label{eq:Rmatrix}
 R\equiv-i\,\hat M_h^{-1/2}\,(\Theta V)^{-1*}\,U_{\rm PMNS}^* \,\hat m_\ell^{1/2}\,\Delta_\ell^{1/2}~.
\end{equation}
This matrix is orthogonal, $R\,R^T=I$. From here, it is straightforward to see that:
\begin{eqnarray}
 \label{eq:Uah}
 U_{ah}&=&i\,U_{\rm PMNS}\, \hat m_\ell^{1/2} \,(\Delta^\dagger_\ell)^{1/2}\,R^\dagger\, \hat M_h^{-1/2} \\
 \label{eq:Usl}
 U_{s\ell} &=&i\,V\, \hat M_h^{-1/2}\,R\,(\Delta_\ell)^{1/2}\, \hat m_\ell^{1/2}
\end{eqnarray}
Then, given the six physical $m_\ell$ and $M_h$, as well as the measured parameters on $U_{\rm PMNS}$, one can build the the whole mixing matrix by specifying the $R$, $V$ and $m_{LL}$ matrices. It is well known that the $R$ matrix can be used to enhance the active-heavy mixing and sterile-light mixing, leading to the possibility of having not-so-heavy $N_h$ with relatively large couplings to the SM. However, this enhancement cannot be arbitrarily large, as the requirement of small $\Theta$ means that $\hat m_\ell^{1/2}(\Delta_\ell^\dagger)^{1/2}\,R^\dagger \hat M_h^{-1/2}$ must also be small.

The $\Delta_\ell$ matrix measures the role of the Weinberg operator in the determination of the $\nu_\ell$ masses, being independent of the heavy/sterile sector. For $\Delta_\ell=I$, the Weinberg operator gives a negligible contribution, while for $\Delta_\ell=0$ it is dominant. If the absolute value of the entries in $\Delta_\ell$ are large, this means that the Seesaw contribution and the Weinberg operator are cancelling each other, indicating the presence of fine-tuning.

Another possible fine-tuning can be found in $m_{RR}$. One would expect the $N_h$ masses to be of the order of the largest of the two contributions to this parameter, see Eq.~(\ref{eq:defmLLmRR}). If this is not the case, it would suggest the presence of a cancellation between them.

In order to avoid this fine-tuning, we require:
\begin{align}
 \frac{v^2\,\alpha_W}{\Lambda}\lesssim m_\ell & & \frac{v^2\,\alpha_{N\phi}}{\Lambda}\lesssim M_h
\end{align}
This means that $\alpha_W/\Lambda$ must be smaller than $\ord{10^{-14}}\,$GeV$^{-1}$. Moreover, for a 10~GeV heavy neutrino, we have $\alpha_{N\phi}/\Lambda\lesssim\ord{10^{-4}}\,$GeV$^{-1}$, with the bound weakening for larger masses. Given the fine-tuning constraint on $\alpha_W/\Lambda$, it is undesirable for all effective operators to share a common origin, as in that case they would be expected to be of a similar order of magnitude, and thus too small to be probed. Now, as discussed in~\cite{Caputo:2017pit}, the presence of a lepton-number symmetry forbids the Weinberg operator, and forces the diagonal terms of the Anisimov-Graesser operator to vanish\footnote{This is valid for two heavy neutrinos. For three heavy neutrinos, two opposite off-diagonals in $\alpha_{N\Phi}$ and $\alpha_{NB}$ would have to vanish as well.}, but puts no constraints on the rest of the coefficients. The breaking of the symmetry would suggest the following hierarchy between them:
\begin{equation}
\label{eq:hierarchy}
\alpha_W\sim\alpha^{\rm diag}_{N\phi}\ll\alpha^{\rm off-diag}_{N\phi}\sim\alpha_{NB}
\end{equation}

These considerations motivate neglecting the contribution of the Weinberg operator to neutrino masses. In other words, in what follows we will diagonalize the mass matrix in Eq.~(\ref{eq:numassmatrix}), using the parametrization of~\cite{Donini:2012tt,Gago:2015vma,Jones-Perez:2019plk}\footnote{For practical purposes, this is equivalent to using Eqs.~(\ref{eq:Uah}) and (\ref{eq:Usl}) with $\Delta_\ell=I$. The advantage of~\cite{Donini:2012tt} is that the parametrization also includes $H$ and $\bar H$ matrices, that keep the mixing unitary when the enhancement factor are too large.}.

It should be noticed that the Anisimov-Graesser and dipole operators generally mix under renormalization group equations with the Weinberg operator, bounded by light neutrino masses. In~\cite{Caputo:2017pit}, an example of one-loop diagram contribution from the Anisimov-Graesser to the Weinberg operator is calculated, giving a rather mild constraint which leaves freedom to assume the above hierarchy.

\section{Couplings and Decay Rates}
\label{sec:Couplings}

\begingroup
\allowdisplaybreaks
When including the effective operators, the coupling terms of the Lagrangian become:
\begin{eqnarray}
 \mathcal L_W&=&\frac{g}{\sqrt2} W_\mu^-\bar\ell_a\gamma^\mu\, U_{a i}\,P_L\, n_i+h.c. \\
 \mathcal L_Z &=&\frac{g}{4c_W} Z_\mu \bar n_i\gamma^\mu\, \left(C_{ij}\,P_L-C_{ij}^*\,P_R\right)\, n_j \nonumber \\
 &&-\frac{s_W}{\Lambda}(\partial_\mu Z_\nu-\partial_\nu Z_\mu)\,\bar n_i\,\sigma^{\mu\nu}\left[(\alpha'_{NB})_{ij}P_L-(\alpha^{\prime\,*}_{NB})_{ij}P_R\right]n_j   \label{eq:L_Znn}\\
 \mathcal L_\gamma &=&\frac{c_W}{\Lambda}(\partial_\mu A_\nu-\partial_\nu A_\mu)\,\bar n_i\,\sigma^{\mu\nu}\left[(\alpha'_{NB})_{ij}P_L-(\alpha^{\prime\,*}_{NB})_{ij}P_R\right]n_j \label{eq:L_gammann}\\
 \mathcal L_h &=&
-\frac{1}{v}h\, \bar n_i\left[\frac{1}{2}\left(C_{ij}\,m_{n_j}+C^*_{ij}\,m_{n_i}\right)-\frac{v^2}{\Lambda}(\alpha^{\prime*}_{N\phi})_{ij}\right]P_R\,n_j \nonumber \\
&&-\frac{1}{v}h\, \bar n_i\left[\frac{1}{2}\left(C_{ij}\,m_{n_i}+C^*_{ij}\,m_{n_j}\right)-\frac{v^2}{\Lambda}(\alpha^{\prime}_{N\phi})_{ij}\right]P_L\,n_j \label{eq:L_higgsnn} \\
 \mathcal L_{hh}&=& \frac{1}{2\Lambda}h^2\,\bar n_{i}\left[(\alpha'_{N\phi})_{ij}P_L+(\alpha^{\prime\,*}_{N\phi})_{ij}P_R\right]n_{j}
\end{eqnarray}
where we have neglected the Weinberg operator, and defined the coefficents:
\begin{eqnarray}
 \label{eq:coefficients0}
 C_{ij}&=& U^*_{a i}\,U_{a j} \\
  \label{eq:coefficients2}
 (\alpha'_{N\phi})_{ij}&=&U_{si}\,(\alpha_{N\phi})_{ss'}\,U_{s'j} \\
 \label{eq:coefficients3}
 (\alpha'_{NB})_{ij}&=&U_{si}\,(\alpha_{NB})_{ss'}\,U_{s'j} 
\end{eqnarray}
\endgroup

In the standard Seesaw, the mixing allows the $N_h$ to interact via the couplings of the active states. In our case, in addition, the mixing allows the $\nu_\ell$ to interact via the effective couplings of the sterile states. This gives rise to new processes, originally forbidden or suppressed in either the Standard Model or Seesaw. We will review how these constrain the effective operators in Section~\ref{sec:Constraints}.

Calculating the lifetime of the heavy neutrinos is essential to determine their observability. For example, a very long-lived $N_h$ could be detector-stable, escaping laboratory experiments as missing energy. On the standard Seesaw, the lifetime of heavy neutrinos with masses in the GeV range can be calculated through three-body decays. These involve a virtual $W$ boson leading to final states with either $l_a\,l_{a'}\,\nu_{a'}$ or $l_a\, q\,q'$ states, or a virtual $Z$ boson implying final $\nu_a\,l_{a'}\,l_{a'}$, $3\,\nu$, or $\nu\,q\,\bar q$ states. These calculations are usually performed via a four-fermion operator, with the gauge bosons integrated out~\cite{Atre:2009rg,Helo:2010cw,Bondarenko:2018ptm,Coloma:2020lgy}, which is appropriate for masses up to around 20 GeV~\cite{Kovalenko:2009td}. 

For heavy neutrinos with masses below the Higgs mass, the Anisimov-Graesser operator plays no role in the determination of the lifetime. In contrast, it is well known that when including the dipole operator a new decay channel into on-shell $\nu\,\gamma$ is opened. This was first calculated in~\cite{Aparici:2009fh}, and we confirm their result:
\begin{equation}
 \label{eq:decayintogamma}
 \Gamma(N_h\to\,\nu\,\gamma)=\frac{2}{\pi}\,c_W^2 M_h^3 \sum_\ell\left|\frac{(\alpha'_{NB})_{\ell h}}{\Lambda}\right|^2
\end{equation}
The partial width of this new process is many orders of magnitude larger than those for the standard Seesaw three-body decays. However, an important point that has been ignored in the literature, until recently~\cite{Barducci:2022hll}, is that the effective dipole coupling also modifies the three-body decay channels involving a virtual $Z$, as well as adding new contributions with virtual photons. In order to understand the importance of these effects, we have performed the calculation of all these processes, and indeed find a significant enhancement to them. This means that, although dominant, the two-body decay in Eq.~(\ref{eq:decayintogamma}) might not always determine the heavy neutrino lifetime on its own. In addition, it might not be always accurate to take the branching ratio into $\nu\,\gamma$ equal to unity.

The exact, analytical formulae for $N_h\to\nu\,q\,\bar q$, $N_h\to\nu\,l_a^+\,l_a^-$ and $N_h\to3\nu$ can be found in Appendix~\ref{app:partial_widths}. The aforementioned enhancement occurs due to new contributions mediated by photon exchange, meaning that, to leading order in $\alpha'_{NB}/\Lambda$, only $N_h\to\nu\,q\,\bar q$ and $N_h\to\nu\,l_a^+\,l_a^-$ are significantly increased. For these decays, in the regime where $\alpha'_{NB}/\Lambda$ is much larger than the standard Seesaw couplings, the partial widths can be written:
\begin{equation}
\label{eq:SimplerPartial}
\Gamma(N_h\to\,\nu\,f\,\bar f)\approx
N_c\,(1+\Delta_{\rm QCD})\, \frac{\alpha_{\rm em}\,Q_f^2}{24\pi^2}\Phi(x_f)\, c_{W}^{2}M_h^{3}\sum_\ell\left|\frac{(\alpha'_{NB})_{\ell h}}{\Lambda}\right|^{2}
\end{equation}
where $N_c$ is the number of colours (3 for quarks, 1 for charged leptons), $\alpha_{\rm em}$ is the fine-structure constant, $Q_f$ is the fermion charge, and $\Phi(x_f)$ is a function obtained after integrating over phase space, depending only on the ratio between the charged fermion and heavy neutrino masses, $x_f$. The function $\Delta_{\rm QCD}$ takes into account QCD corrections to decays into quark final states~\cite{Bondarenko:2018ptm}, and vanishes for the purely leptonic decays.

\begin{figure}[tbp]
\centering
\includegraphics[width=0.48\textwidth]{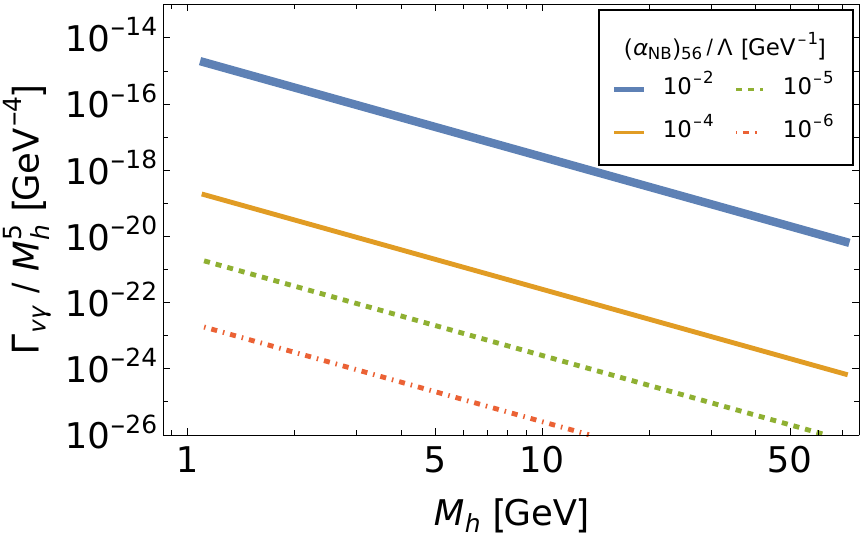} \hfill
\includegraphics[width=0.48\textwidth]{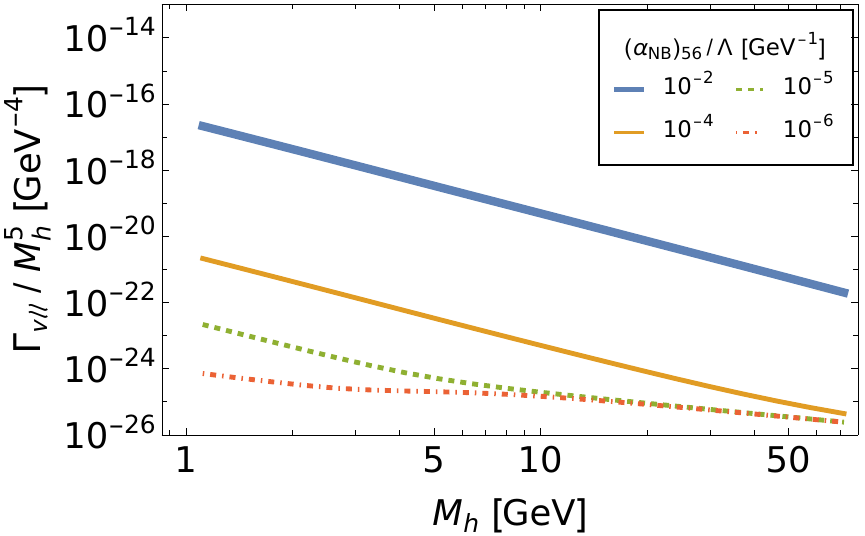} \vspace{3mm} \\
\includegraphics[width=0.48\textwidth]{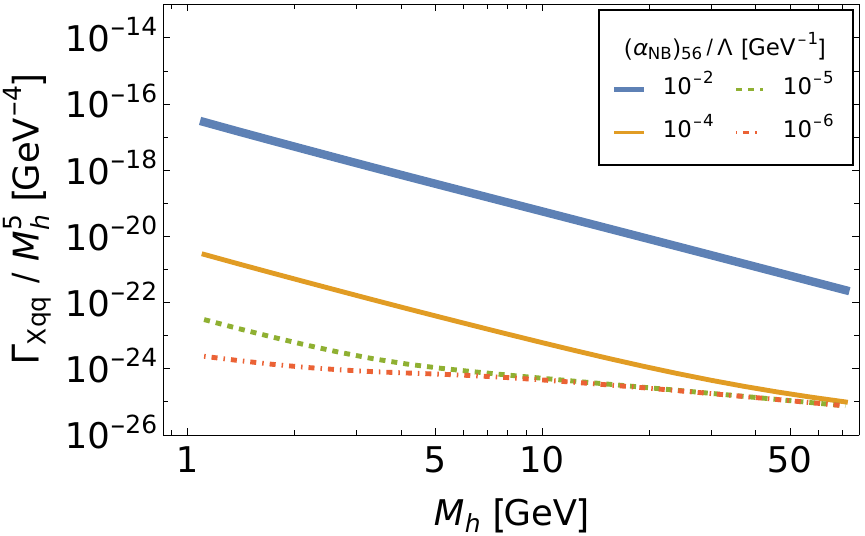} \hfill
\includegraphics[width=0.48\textwidth]{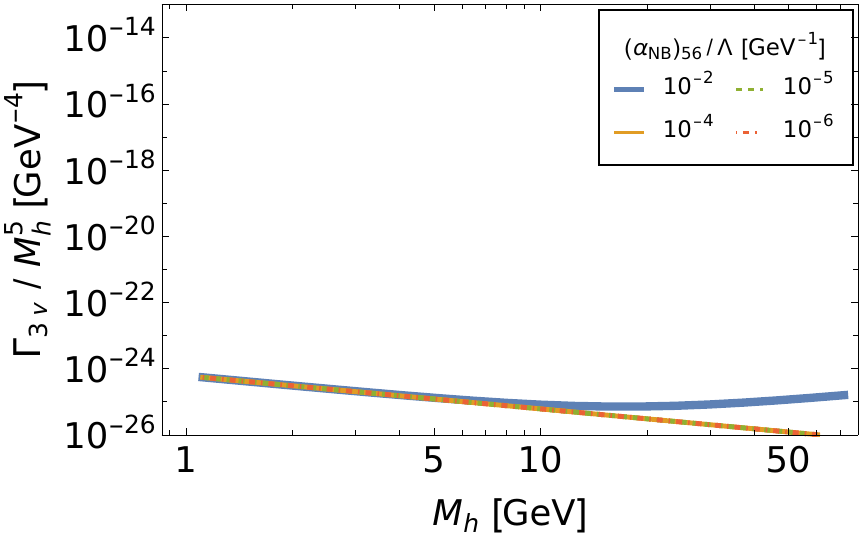}
\caption{Heavy neutrino partial width into several types of final state when including effective operators, normalized with respect to $M_h^5$. From left to right, top to bottom, we show the partial widths for $N_h\to\nu\,\gamma$, $N_h\to\nu\,l_{a^{(\prime)}}l_a$, $N_h\to X\, q \,q'$ and $N_h\to3\,\nu$. On each panel we plot the widths for $(\alpha_{NB})_{56}/\Lambda=10^{-2},\,10^{-4},\,10^{-5},\,10^{-6}$~GeV$^{-1}$ in blue (thick), orange (thin), green (dashed), and red (dot-dashed), respectively. We take minimal mixing.} 
\label{fig:NuDecayWidthsEff}
\end{figure}
In order to present our findings, we take $N_5$ as the lightest heavy neutrino, and show in Figure~\ref{fig:NuDecayWidthsEff} its partial widths for several values of $(\alpha_{NB})_{56}/\Lambda$, taking minimal mixing, i.e.\ $R=I$. We find that the three-body widths for $(\alpha_{NB}/\Lambda)\leq10^{-6}$~GeV$^{-1}$ are indistinguishable from the standard Seesaw, so they can be used as benchmarks for comparison.

In the Figure, we show that the partial widths into charged leptons and quarks can be significantly enhanced. As commented earlier, this is attributed to the dominance of the virtual photon contribution, which is not suppressed by the $Z$ mass. As seen in Eq.~(\ref{eq:SimplerPartial}), the size of these effects are proportional to $M_h^3$. This should be compared with the standard Seesaw processes, which are proportional to $M_h^5$, meaning that at high mass the latter contributions become more relevant. In contrast, the partial width into three light neutrinos is not enhanced, except for very large $\alpha_{NB}/\Lambda$ and masses\footnote{It is important to notice that the photon-dominated three-body decays are not calculated via a four-fermion operator, so they remain accurate for large $M_h$. This is not the case for $N\to3\nu_\ell$, but should not be a cause for concern, given its small impact on the heavy neutrino lifetime.}.

In addition to the three-body decays, we also show the partial width of the aforementioned two-body decay into $\nu\,\gamma$. Comparison with the other partial widths suggests that $\nu\,\gamma$ decay can easily dominate the total width, depending on the heavy neutrino mass and the exact value of the dipole coupling $\alpha_{NB}/\Lambda$.

\begin{figure}[tb]
\centering
\includegraphics[width=0.48\textwidth]{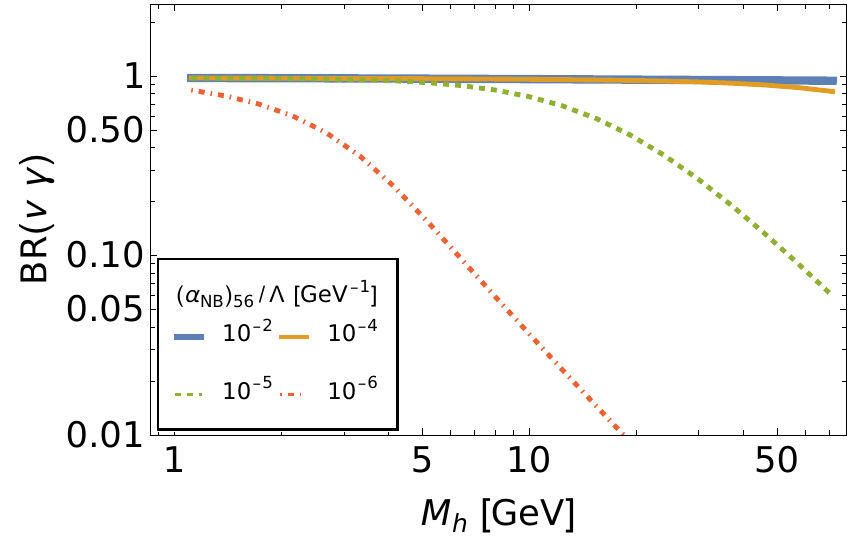} \hfill
\includegraphics[width=0.48\textwidth]{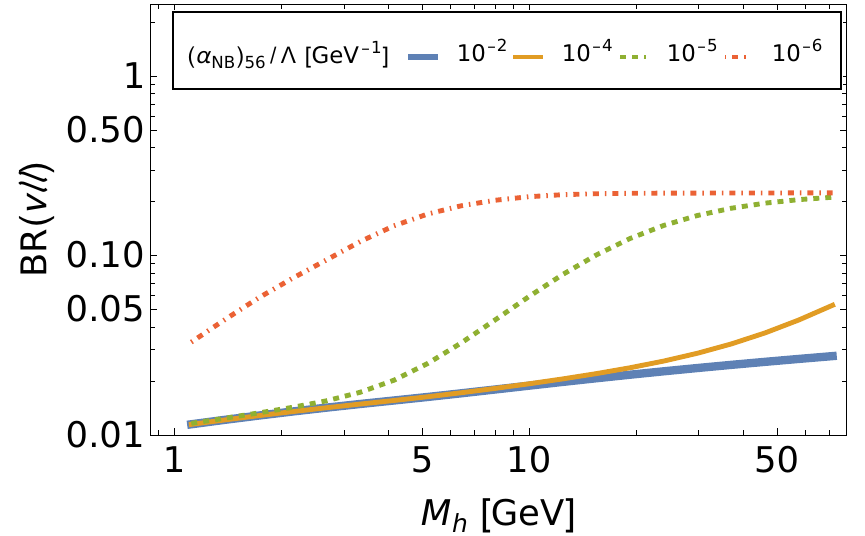} \vspace{3mm} \\
\includegraphics[width=0.48\textwidth]{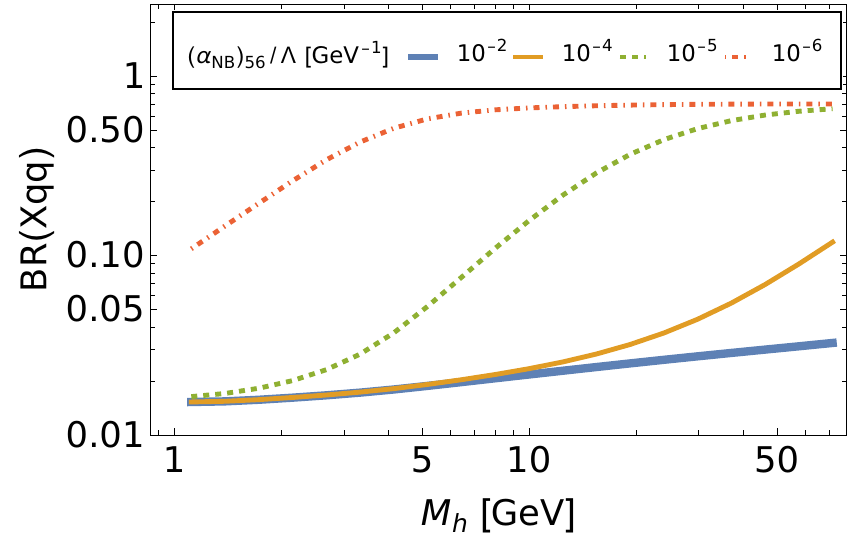}
\hfill
\includegraphics[width=0.48\textwidth]{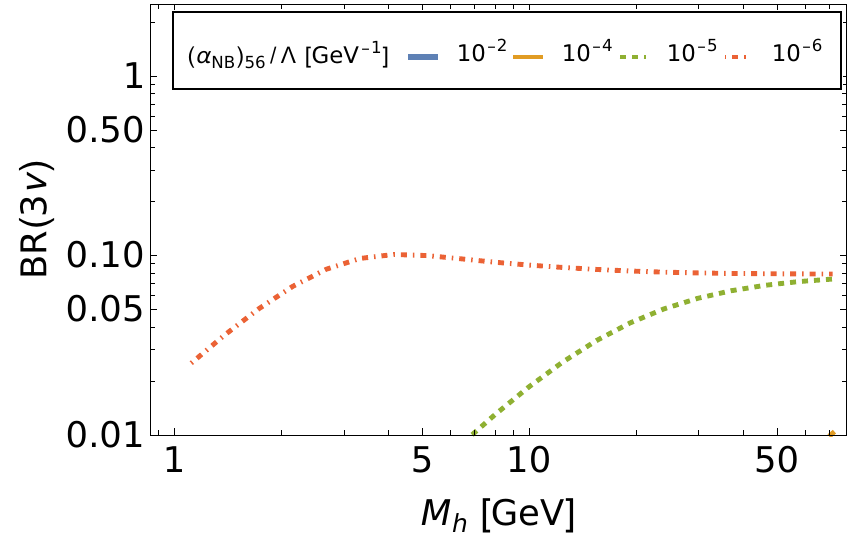} 
\caption{Heavy neutrino branching ratios. Choice of parameters as in Figure~\ref{fig:NuDecayWidthsEff}.} 
\label{fig:Nu3BREff}
\end{figure}
In order to better understand the interplay between the widths, we show in Figure~\ref{fig:Nu3BREff} the branching ratios into the different channels. We would like to emphasize that the branching ratios are for all practical purposes independent of the choice of the $R$ matrix, defined in Eq.~(\ref{eq:Rmatrix}). We find that, for all of the evaluated values of $M_h$, taking $BR(N_h\to\nu\,\gamma)=1$ is a very good approximation when taking $\alpha_{NB}/\Lambda\gtrsim10^{-4}\,$GeV$^{-1}$. Nevertheless, the branching ratio can be much lower. For example, for $\alpha_{NB}/\Lambda=10^{-5}\,$GeV$^{-1}$, the branching ratio is under $50\%$ for $M_h\gtrsim20\,$GeV. If $\alpha_{NB}/\Lambda=10^{-6}\,$GeV$^{-1}$, the same happens if $M_h\gtrsim2.5\,$GeV.

Let us turn now to three-body branching ratios. For low masses, the increase in the $N_h\to\nu\,\gamma$ partial width drives the three-body branching ratios to very low values. Thus, even though the modifications to the three-body widths are larger in this regime, as long as $(\alpha_{NB}/\Lambda)\gtrsim10^{-5}$\,GeV$^{-1}$, the huge contribution from the two-body decays make them less relevant. For large masses the three-body branching ratios increase, being larger for smaller $\alpha_{NB}/\Lambda$. This can be understood by considering that, even though for smaller $\alpha_{NB}/\Lambda$ all contributions from the effective operator decrease, the standard Seesaw processes remain the same, giving a fixed minimum contribution to  three-body decays only.

\begin{figure}[tbp]
\centering
\includegraphics[width=0.49\textwidth]{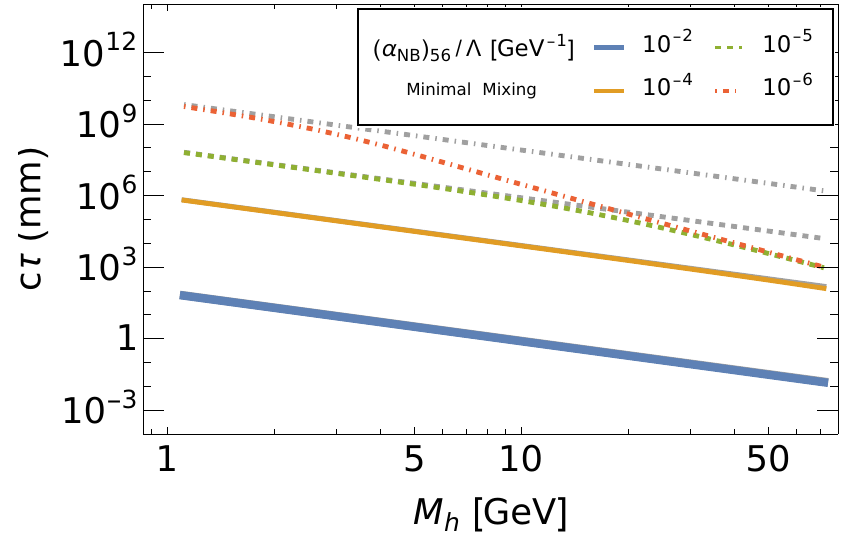} \hfill
\includegraphics[width=0.49\textwidth]{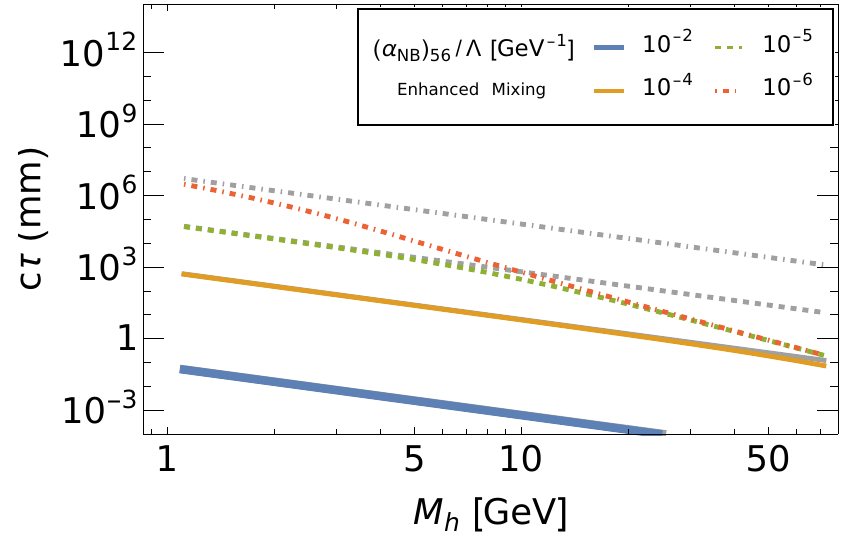} \vspace{3mm} \\
\includegraphics[width=0.49\textwidth]{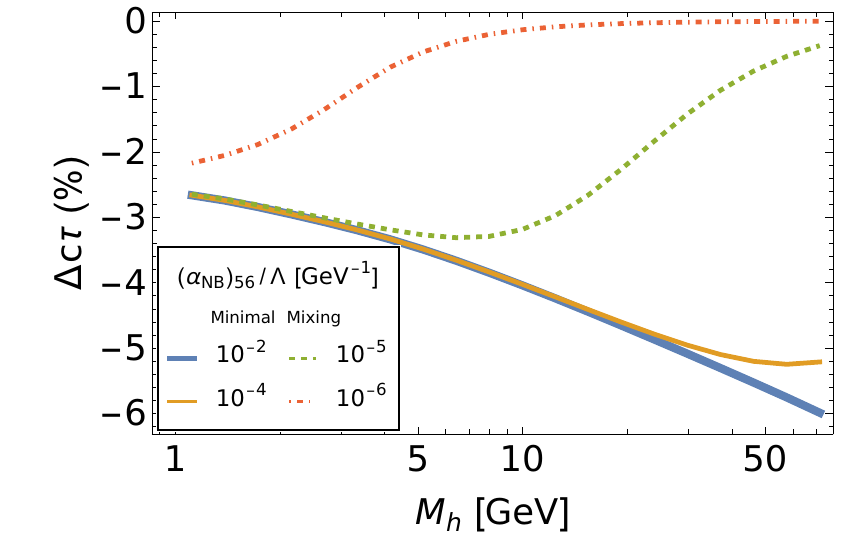} \hfill
\includegraphics[width=0.49\textwidth]{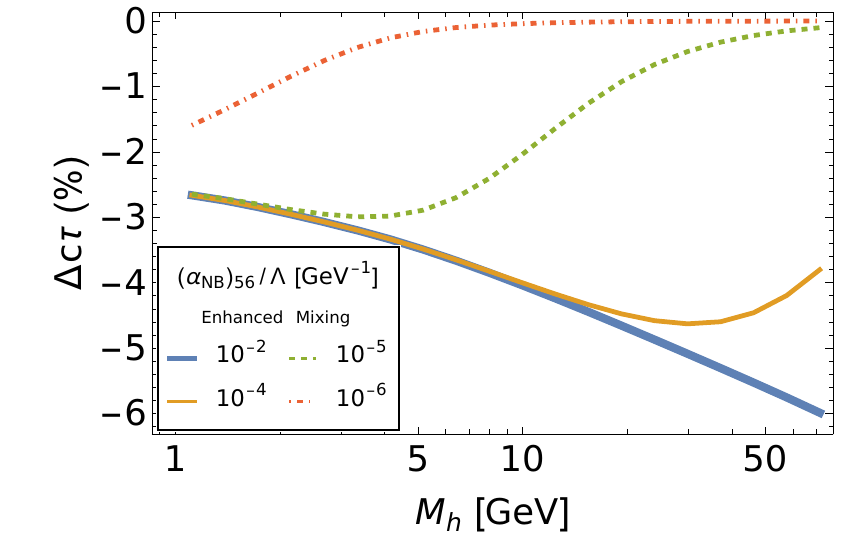}
\caption{Top panels: Heavy neutrino decay length. Gray lines indicate the expected decay length when considering only two-body decays. Bottom panels: Difference in decay length when performing the full calculation, versus including only $N\to\nu\gamma$ and the standard Seesaw three-body decays. Left: Choice of parameters as in Figure~\ref{fig:NuDecayWidthsEff}. Right: The same, but with enhanced sterile-light mixing.} 
\label{fig:NuGammaLifeEff}
\end{figure}
We now analyze the heavy neutrino lifetime. First, on the upper panels of Figure~\ref{fig:NuGammaLifeEff} we report the expected decay length $c\tau$ of the heavy neutrino in presence of the new couplings. On the left panel we show results for minimal mixing, and on the right for a choice of $R$ leading to a $10^3$ enhancement in the sterile-light mixing squared. Within the evaluated mass range, the $N_h$ with minimal mixing always have decay lengths longer than one meter\footnote{A decay length of one meter implies a lifetime around three nanoseconds.} as long as $(\alpha_{NB}/\Lambda)\lesssim10^{-5}$~GeV$^{-1}.$ As seen on the right panel, enhancing the neutrino mixing by taking non-trivial $R$ would decrease the decay length, meaning that heavy neutrinos on this mass range can always be made unstable within collider scales.

On the upper panels, we also show in gray lines the expected decay length if one considers that the width is exclusively determined by Eq.~(\ref{eq:decayintogamma}). Thus, for $\alpha_{NB}/\Lambda\gtrsim10^{-4}\,$GeV$^{-1}$, one can take the lifetime directly from $N_h\to\nu\,\gamma$, without having to include three-body decays. This is not the case for smaller $\alpha_{NB}/\Lambda$, if one does not include the latter, the decay length could be off by many orders of magnitude.

Finally, on the lower panels of Figure~\ref{fig:NuGammaLifeEff}, we assess the impact of the new contributions in Eq.~(\ref{eq:SimplerPartial}) on the decay length by comparing the full calculation with what is obtained by including only $N_h\to\nu\,\gamma$ and the standard Seesaw three-body decays. We see that, naturally, for a fixed mass the deviation increases with $(\alpha_{NB})/\Lambda$. Their relevance with respect to $M_h$ depends on their interplay with the standard Seesaw contributions, which in turn depends on $R$. However, in all cases the inclusion of these new effects leads to a reduction in the decay length of a few percent, suggesting that the need of including them in calculations can be delayed to a precision era, after a putative discovery.

\section{Constraints on Effective Dipole Operator}
\label{sec:Constraints}

The partial widths and lifetimes shown in the previous Section are very sensitive to the value of  $\alpha_{NB}/\Lambda$. It is thus necessary to review current constraints on this coupling. Since we are interested in long-lived heavy neutrinos detectable at the LHC, we will pay particular attention to bounds relevant for GeV-scale masses.

When considering vertices involving two light neutrinos and a photon, one finds that the effective coupling induces transition magnetic and electric dipole moments, $\mu_\nu$ and $\epsilon_\nu$~\cite{Giunti:2014ixa}:
\begin{align}
(\mu_\nu)_{\ell\ell'}=4i\,c_W\,\frac{\Im m(\alpha'_{NB})_{\ell\ell'}}{\Lambda} & &
(\epsilon_\nu)_{\ell\ell'}=4i\,c_W\,\frac{\Re e(\alpha'_{NB})_{\ell\ell'}}{\Lambda}~,
\end{align}
where $\alpha'_{NB}$ is defined in Eq.~(\ref{eq:coefficients3}). These are forbidden at tree level in both the SM and standard Seesaw, and are subject to constraints. For instance, both magnetic and electric dipole moments of light neutrinos can affect the electron energy spectrum in $\nu-e$ scattering, which can be probed in solar, reactor and accelerator experiments~\cite{Giunti:2014ixa,Giunti:2015gga}. In particular, the Borexino experiment has placed a bound on $\mu_\nu^{eff}<2.8\times10^{-11}\,\mu_B$~\cite{Borexino:2017fbd}. An even stronger bound comes from the brightness of red giant stars, where the transition dipole moments enable plasmon decay, $\gamma^*\to\nu\nu$, and thus allows energy to be released from the star. Observations from the galactic globular cluster $\omega$ Centauri constrain $\sqrt{\sum_{\ell\ell'}(|(\mu_\nu)_{\ell\ell'}|^2+|(\epsilon_\nu)_{\ell\ell'}|^2)}\lesssim1.2\times10^{-12}\,\mu_B$~\cite{Capozzi:2020cbu}.

In terms of the effective operator, light neutrino dipole moments give us the bound:
\begin{eqnarray}
 4\,c_W\left|U_{s\ell}\frac{(\alpha_{NB})_{ss'}}{\Lambda}U_{s'\ell'}\right|\lesssim3\times10^{-10}\,{\rm GeV}^{-1}.
\end{eqnarray}
From Equation~(\ref{eq:Usl}), if we take $R=I$ we have $U_{s\ell}\sim\sqrt{m_\ell/M_h}$, such that if light neutrinos have masses below $10^{-2}$~eV, these bounds on $\alpha_{NB}/\Lambda$ essentially vanish for heavy neutrino masses in the 100 MeV scale, or larger. In principle, it could be expected that by taking non-trivial $R$ one could enhance the dipole moments. However, when considering the structure of $U_{s\ell}$ shown in Eq.~(\ref{eq:Usl}), along with the antisymmetry of $\alpha_{NB}$, one finds that the leading contributions always cancel. Thus, in the following we shall not take these bounds into account.

The dipole moments of heavy neutrinos themselves can be constrained using atomic spectroscopy. In~\cite{Bolton:2020xsm}, the long-range potentials from $N_h$ exchange are calculated, establishing a limit of $4\,c_W\,\alpha_{NB}/\Lambda\lesssim10^{-2}\,\mu_B$, for a specific benchmark.

On can find further constraints when considering transition moments involving one light and one heavy neutrino. In addition to $\nu-e$ scattering and plasmon decay in red giants, the authors in~\cite{Brdar:2020quo} have considered the impact on Supernova neutrino emission, $^4$He abundance from Big Bang nucleosynthesis, and the $N_{\rm eff}$ measurement from the cosmic microwave background. Although strong, all of these bounds vanish for $N_h$ masses larger than about 300~MeV.

High energy experiments can also probe the latter transition dipole moments. The most important constraints for heavy neutrinos in the GeV scale come from LEP searches for photons and missing energy~\cite{L3:1992cmn,OPAL:1994kgw,DELPHI:1996drf}, and single $\gamma$ production at NOMAD~\cite{Gninenko:1998nn,NOMAD:1997pcg}. For example, the authors in~\cite{Aparici:2009fh} surveyed the possible LEP constraints on its effective coupling $\alpha_{NB}/\Lambda$ from the $Z$ invisible width and $Z$ decays into photons and missing energy. They considered $e^+e^-\to N_5\,N_6$, with $N_6\to N_5\,\gamma$ and both $N_5$ escaping the detector. They found that, if $N_5$ is massless and $N_6$ is lighter than the Z, they ruled out $(\alpha_{NB})_{56}/\Lambda>2.5\times10^{-5}$ GeV$^{-1}$.

Another example can be found in~\cite{Magill:2018jla}, where the authors put several constraints on the dimension-6 $N_h$ dipole coupling to a $\nu_\ell$ and a photon. For the GeV scale, their most important constraint comes from $e^+e^-\to N_h\,\nu_\ell$ at LEP, where they exclude $\alpha'_{NB}/\Lambda\gtrsim2\times10^{-5}$~GeV$^{-1}$. Since this analysis considers that the light and heavy neutrinos couple exclusively via a Dirac-type dipole interaction, we have considered pertinent to reformulate their bounds in our context. To this end, we have calculated the $e^+e^-\to N_h\,\nu_\ell$ cross section analytically, including both $Z$ and photon exchange, at $s=m_Z^2$. In contrast to~\cite{Magill:2018jla}, we have included both $(\alpha'_{NB})_{\ell h}/\Lambda$ and $C_{\ell h}$ couplings to the $Z$ in Eq.~(\ref{eq:L_Znn}), and consider $N_h\to\nu\,\gamma$ branching ratios different from unity. If one neglects the electron mass, and does not impose any cuts, the total production cross section is:
\begin{multline}\label{eq:SigLep}
\sigma_{N\nu}=\frac{(M_h^2-m_Z^2)^2}{2\pi m_Z^2\Gamma_Z^2}(c_V^2+c_A^2)\bigg\{
\left|\frac{(\alpha'_{NB})_{\ell h}}{\Lambda}\right|^2
\frac{e^2(2M_h^2+m_Z^2)}{3 c_W^2 m_Z^2}\left(1+\frac{4c_W^2\Gamma_Z^2}{(c_V^2+c_A^2)m_Z^2}\right) \\
+\frac{1}{6}|C_{\ell h}|^2 G_F^2(M_h^2+2m_Z^2)\,c_W^2
-\sqrt2\,\Re e\left[\frac{(\alpha'_{NB})_{\ell h}}{\Lambda}C_{\ell h}\right] e\, G_F\, M_h
\bigg\}
\end{multline}
By neglecting $C_{\ell h}$, one can reproduce the cross-section reported in~\cite{Magill:2018jla}\footnote{Notice that the $d_{\gamma, Z}$ coupling in~\cite{Magill:2018jla} corresponds to $2\,c_W\,\alpha'_{NB}/\Lambda$.}.

For comparison purposes, we apply energy and angular cuts on the final photon, and require the cross-section not to exceed $0.1\,$pb \cite{Lopez:1996ey}. The photon energy $E_\gamma$ must be above $0.7\,$GeV, and the photon angle $\theta$ with respect to the beamline must satisfy $|\cos\theta|\leq0.7$. We require the heavy neutrino to decay within two meters of the interaction point. Technical details of the calculation can be found in Appendix~\ref{app:LEP}

\begin{figure}[tbp]
\centering
\includegraphics[width=0.49\textwidth]{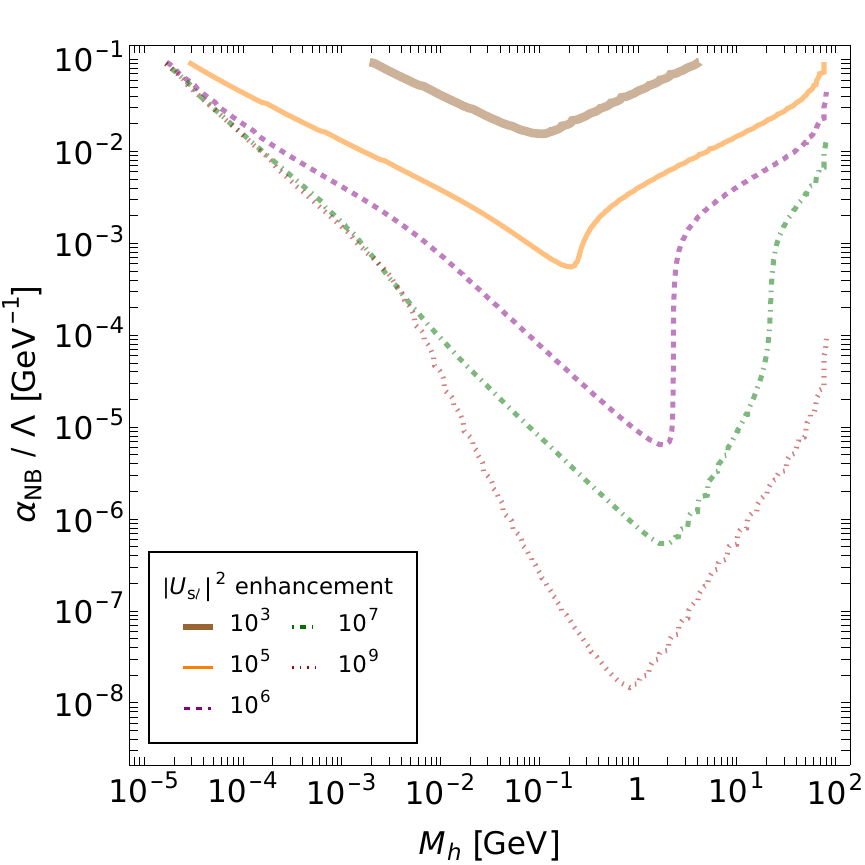} \hfill
\includegraphics[width=0.49\textwidth]{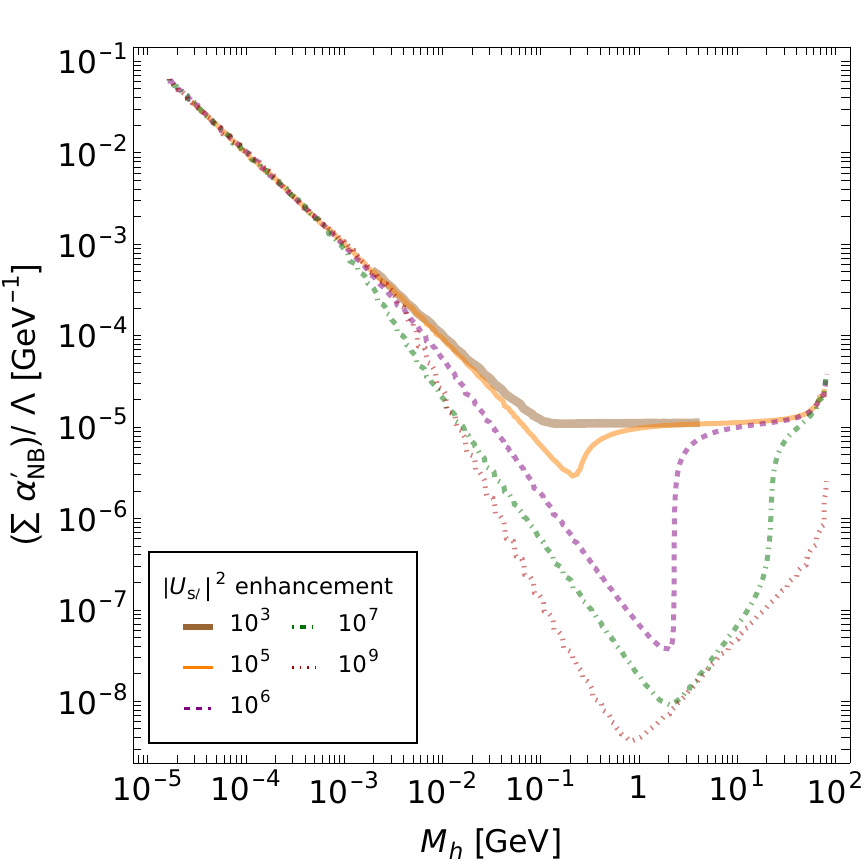}
\caption{LEP bounds on $(\alpha_{NB})_{56}/\Lambda$. On all plots, we show the bound for $|U_{s\ell}|^2$ enhancements of $10^3$, $10^5$, $10^6$, $10^7$ and $10^9$ in brown (thick), orange (thin), purple (dashed), green (dot-dashed) and red (dotted) lines, respectively. On the left panel we show the bound on the parameter itself, while on the right panel we display the bound on $\sum(\alpha'_{NB})/\Lambda$.} 
\label{fig:LepBound}
\end{figure}
Our bounds are shown for $N_5$ in Fig.~\ref{fig:LepBound}, for several enhancements of $|U_{s\ell}|^2$ from $R$. We show the exclusion regions in terms of both $(\alpha_{NB})_{56}$ and $(\alpha'_{NB})_{\ell5}$. Strictly speaking, since the heavy neutrino hadronic partial widths are calculated via three-body decays, our results are accurate only for masses above 1 GeV. However, we extend our plot to lower masses for a better comparison with~\cite{Magill:2018jla}.

Let us focus first on the left panel, which shows the constraints in terms of $(\alpha_{NB})_{56}/\Lambda$. As expected, the larger the enhancement in $|U_{s\ell}|^2$, the stronger the bound on this parameter. If there is no enhancement, we find that LEP cannot place any constraints. For small enhancement ($10^3$), the exclusion region has a triangular shape, with the right slope being due to the $\sqrt{m_\ell/M_h}$ suppression of $U_{s\ell}$ in $\sigma_{N\nu}$, and the left slope being due to the heavy neutrino escaping the detector. As the enhancement grows ($10^5$), the bound acquires a bump in the middle, which is attributed to an increased contribution to $\sigma_{N\nu}$ from the $C_{\ell h}$ terms. The size of the bump increases with the enhancement ($10^6$), until it eventually receives a second cut on the right, from a decreased $N\to\nu\,\gamma$ branching ratio ($10^7$), as can be seen in Fig~\ref{fig:Nu3BREff}. The exclusion region can be expanded up to an enhancement around $10^9$, after which the $\Theta$ matrix in Eq~(\ref{eq:ThetaDiag}) cannot be considered small anymore. A careful analysis using the parametrization of~\cite{Donini:2012tt} shows that the unitarity of the mixing matrix does not allow further enhancement. Thus, the strongest constraint on $(\alpha_{NB})_{56}/\Lambda$ is of $1.4\times10^{-8}$ GeV$^{-1}$, for masses around 800~MeV.

In order to properly contrast our model with that of~\cite{Magill:2018jla}, we show on the right panel the same information in terms of $\sum\alpha'_{NB}\equiv\sqrt{\sum_{\ell=1}^3{|(\alpha'_{NB})_{\ell 5}|^2}}$. When taking small enhancement ($10^3$), the exclusion region is flat down to masses of around 100~MeV, below which the probability the heavy neutrino escapes the detector becomes non-negligible, and the bound weakens. For large masses, the bound is flat since the $\sqrt{m_\ell/M_h}$ suppression of $U_{sl}$ is absorbed within $\alpha'_{NB}$. As the enhancement from $R$ grows, we find a behaviour similar to that for the $\alpha_{NB}/\Lambda$ constraints. In all cases, the bound is weakened at small masses, due to the cut on the heavy neutrino decay length.

An important point to take into account is that, in Figure 9 of~\cite{Magill:2018jla}, the aforementioned flat exclusion region is extended down to 1~MeV masses, instead of 100~MeV. We have not been able to reproduce this result, in fact, our analysis was expected to produce stronger bounds, given that the heavy neutrinos in this model have shorter lifetimes, and we are accepting longer decay lengths.

\section{Searches for Non-Pointing Photons at the LHC}
\label{sec:NonPointing}

We now turn to possible LHC probes of the model, which can be sensitive to GeV-scale heavy neutrinos. Here, current bounds on heavy neutrinos~\cite{CMS:2022rqc, Sirunyan:2018mtv, ATLAS:2019kpx, CMS:2019lwf, ATLAS:2018dcj} are based on standard Seesaw interactions, and can be avoided if the latter decays exclusively into final states with photons. It is then of interest to find out if LHC searches can probe our model at all, and test the existence of the dipole coupling.

With this in mind, we identify two regimes: one where the $N_h$ decays promptly, and one where it is a long-lived particle. For this work, we are interested in the case where the heavy neutrinos are long-lived, meaning that we focus on searches for particles with large lifetimes decaying into displaced photons, such as~\cite{ATLAS:2014kbb, ATLAS:2013etx, CMS:2019zxa, CMS:2012bbi, Mahon:2021bai, Zhang:2020awq}.

As seen in Figure~\ref{fig:NuGammaLifeEff}, the $N_h$ decay length depends on the Seesaw mixing, such that its enhancement by $R$ leads to shorter lifetimes. This means that if one wants to have a specific decay length, having an enhancement requires smaller $\alpha_{NB}$. On the other hand, the $N_h\to\nu\,\gamma$ branching ratio depends only on the mass and value of $\alpha_{NB}$, so, for a given mass, in order to maximize the branching ratio into photons, we need the largest possible $\alpha_{NB}$. From this, we conclude that searches for displaced photons will be most sensitive to the case where there is no $R$ enhancement on the Seesaw mixing, which is precisely the regime where LEP has no sensitivity. In other words, these kind of searches are complementary to those at LEP.

Since we focus on scenarios with no enhancement from $R$, all of the $N_h$ production modes at colliders from the standard Seesaw~\cite{Deppisch:2015qwa,Antusch:2016ejd,Abdullahi:2022jlv} are strongly suppressed. In light of this, the best bet for heavy neutrino production comes from the Anisimov-Graesser operator, as can be seen in Eq.~(\ref{eq:L_higgsnn}). This leads to the following partial width for exotic Higgs decay into two heavy neutrinos~\cite{Graesser:2007yj,Aparici:2009fh}:
\begin{multline}
\Gamma(H\to N_h\,N_{h'}) =\\
S_{hh'}\frac{v^2}{2\pi}
\frac{\sqrt{\lambda(m_H^2,\,M_h^2,\,M_{h'}^2)}}{m_H^3}
\left|\frac{(\alpha_{N\phi})_{hh'}}{\Lambda}\right|^2
\left(m_H^2-M_h^2-M_{h'}^2-2M_h\,M_{h'}\cos2\delta_{hh'}\right)
\end{multline}
where $m_H$ is the Higgs mass, $\delta_{hh'}=\arg[(\alpha_{N\phi})_{hh'}]$ and the $\lambda$ function is defined in Appendix~\ref{app:partial_widths}. The factor $S_{hh'}=1/2$ if $h=h'$, and is equal to unity otherwise.

It is important to emphasize that this production mode is driven by $\alpha_{N\phi}$ entirely, while the heavy neutrino $\nu\,\gamma$ decay is dominated by the $\alpha_{NB}$ dipole coupling. Assuming that the value of $\alpha_{NB}$ is such that the $N_h$ decays within the detector, any experimental constraints coming from displaced photon searches would lead to bounds on the $H\to N_h\,N_{h'}$ branching ratio, effectively constraining $\alpha_{N\phi}$. Notice that any excess in such a search would also imply a lower bound on $\alpha_{NB}$, as a too small dipole coupling would lead to a detector stable heavy neutrino.

Given the arguments at the end of Section~\ref{sec:massparam}, we expect $\alpha_{N\phi}$ to have vanishing diagonal entries. Thus, in the following, we consider the decay $H\to N_5\,N_6$, taking both heavy neutrinos as mass degenerate. Such a degeneracy is to be expected by the lepton-number symmetry justifying the hierarchy in Eq.~(\ref{eq:hierarchy})~\cite{Shaposhnikov:2006nn,Kersten:2007vk,Lopez-Pavon:2012yda,Antusch:2017ebe,Hernandez:2018cgc}. One can show that, in addition, when the masses are degenerate, then the $U_{sl}$ mixing for both neutrinos becomes of the same order of magnitude. This should increase the $\sigma_{N\nu}$ cross-section at LEP in Eq.~(\ref{eq:SigLep}) by a factor two, but will not exclude any of the benchmark points we will use.

Non-standard decays of the Higgs have been already constrained by both ATLAS and CMS collaborations~\cite{CMS:2018uag,ATLAS:2019nkf}, determining that the Higgs branching ratio to exotic states must not exceed $0.21$ at $95\%$~C.L.\footnote{Notice that if the heavy neutrinos were detector-stable, and thus invisible, the limits on the branching ratio are between $11\%$~\cite{ATLAS-CONF-2020-052} and $18\%$~\cite{CMS:2022qva}.} In order to maximize the sensitivity of this search to our model, we fix $(\alpha_{N\phi})_{56}/\Lambda$ such that this bound is saturated. The masses and effective couplings we will use can be seen in Table~\ref{Table:Benchmarks}. We have checked that the additional contribution to the $N_h$ mass is at most $20\%$ of the tree level value, so we have no fine-tuning.

\begin{table}
\begin{center}
\begin{tabular}{|C{5cm}||C{2cm}|C{2cm}|C{2cm}|}
\hline
$M_h$ [GeV] & 10 & 30 & 50  \\
\hline 
$(\alpha_{N\phi})_{56}/\Lambda$ [GeV$^{-1}$] & $3.0\times10^{-5}$ & $3.6\times10^{-5}$ & $6.4\times10^{-5}$  \\
\hline 
$(\alpha_{NB})_{56}/\Lambda$ [GeV$^{-1}$] (2014) & $6.5\times10^{-4}$ & $1.4\times10^{-4}$ & $4.8\times10^{-5}$  \\
\hline 
$(\alpha_{NB})_{56}/\Lambda$ [GeV$^{-1}$] (2021) & $7.9\times10^{-4}$ & $1.5\times10^{-4}$ & $6.3\times10^{-5}$  \\
\hline 
\end{tabular}
\end{center}
\caption{\label{Table:Benchmarks} Benchmarks used in our analysis. In the second row we show the effective heavy neutrino coupling to the Higgs $\alpha_{N\phi}/\Lambda$ giving a $H\to N_5 N_6$ branching ratio of $21\%$. The third and fourth rows give the value of the dipole couplings $\alpha_{NB}/\Lambda$ optimal for the searches outlined in Sections~\ref{sec:Paper} and~\ref{sec:Thesis}, respectively.}
\end{table}
In order to calculate the expected number of $N_5\,N_6$ pairs at the LHC, we consider Higgs production via gluon fusion (GF) and vector boson fusion (VBF) processes. These are the two most important interactions for Higgs production at the LHC~\cite{Djouadi:2005gi,Dittmaier:2011ti}. The signal events were generated as described in Appendix~\ref{app:NumericalTools}.

\subsection{8 TeV ATLAS Search (2014)}
\label{sec:Paper}

Here we first follow the analysis performed in~\cite{ATLAS:2014kbb}, a search for delayed non-pointing photons in the diphoton plus missing transverse momentum (MET) final state, in $8$ TeV collisions from ATLAS. As far as we are concerned, this work is the latest published paper by an LHC experiment featuring non-pointing photons.

The search uses the full 20.3~fb$^{-1}$ data sample from 2012, exploiting the capabilities of the ATLAS electromagnetic calorimeter (ECAL) to measure the flight direction as well as the time of flight of photons. An electromagnetic shower produced by a photon is precisely measured in three layers of different depth, with varying lateral segmentation, allowing the reconstruction of the flight direction of the photon. From this one can determine the pointing variable $|\Delta z_{\gamma}|$, defined as the separation between the extrapolated origin of the photon and the position of the primary vertex of the event, measured along the beamline. In addition, ATLAS can also measure the relative arrival time $t_{\gamma}$ of the photon to the calorimeter, compared to that expected for a prompt photon from the hard collision. Then, for prompt decays, both $|\Delta z_\gamma|$ and $t_\gamma$ are naively expected to be zero, meaning these are useful handles for identifying neutral long-lived particles decaying into photons.

The original search targeted a gauge mediated supersymmetry breaking (GMSB) signal where the next to lightest SUSY particle (NLSP) is a neutralino, which can decay to a photon and a stable gravitino. Neutralino pair production would lead to a $\gamma \gamma + {\rm MET}+X$ final state, featuring delayed, non-pointing photons. The selected events were collected by an online trigger requiring at least two ``loose" photons with $|\eta|<2.5$, one with $E_{T}>35$~GeV and another with $E_{T}>25$~GeV. The offline selection then requires two loose photons with $E_{T}>50$~GeV and $|\eta|<2.37$ (excluding the transition region between the barrel and endcap at $1.37<|\eta|< 1.52$), with at least one of them in the barrel region ($|\eta|<1.37$). The isolation criteria demanded the transverse energy deposit in a cone with $\Delta R= 0.4$ around each photon to be less than $4$~GeV. 

\begin{table}
\begin{center}
\begin{tabular}{|c||C{2cm}|C{2cm}|C{2cm}|C{2cm}|}
\hline
\multicolumn{5}{|c|}{Region definition, based on MET [GeV]} \\
\hline
Analysis & BKG & CR1 & CR2 & SR \\
\hline \hline
8 TeV (2014) \cite{ATLAS:2014kbb} & $< 20$ & $[20,\,50]$ & $[50,\,75]$ & $>75$ \\
\hline 
13 TeV (2021) \cite{Mahon:2021bai} & $<30$ & \multicolumn{2}{c|}{$[30,\,50]$} & $>50$ \\
\hline
\end{tabular}
\end{center}
\caption{\label{Table:Regions1} Background (BKG), control (CR) and signal regions (SR) for non-pointing photon searches.}
\end{table}
The selected diphoton sample is divided into exclusive subsamples according to the value of MET: Background Region (BKG), two Control Regions (CR1 and CR2), and Signal Region (SR) (see Table~\ref{Table:Regions1}). The control regions are used to validate the analysis technique and background modelling. Within the signal region, the calculation of $|\Delta z_\gamma|$ and $t_\gamma$ is carried out only with the photon with the maximum $t_\gamma$, in the barrel region. The analysis finally divides the sample into six exclusive categories according to the value of $|\Delta z_{\gamma}|$, and then shows the $t_{\gamma}$ distributions of each of the categories to determine possible signal contributions. The binning in $|\Delta z_{\gamma}|$ and $t_{\gamma}$ is chosen to optimize the expected sensitivity to the GMSB signal. With this, the analysis in~\cite{ATLAS:2014kbb} excludes a certain range of neutralino lifetimes for masses lighter than 440 GeV (see Figure 7 in the reference).

In order to implement this search, we generated $N_5,\,N_6$ pairs and allowed them to decay into photons. The $\texttt{HepMC}$ data was extracted, upon which the aforementioned cuts were applied. We calculated $|\Delta z_\gamma|$ using:
\begin{eqnarray}
|\Delta z_\gamma|=
\left|\frac{r_z-p_z(\vec p\cdot\vec r)/|\vec p\,|^2}{1-p_z^2/|\vec p\,|^2}-z_{\rm PV}\right|
\end{eqnarray}
Here, $\vec r$ denotes the position where the heavy neutrino decays, and $\vec p$ is the momentum of the daughter photon. The variables $r_z$ and $p_z$ denote the components of the former on the $z$-axis. In addition, $z_{PV}$ is the position of the primary vertex along the beamline. The other important variable in the analysis is the relative arrival time $t_\gamma$. In order to obtain it, we first calculated the time $t_0$ a prompt photon would take to reach the ATLAS ECAL, as a function of the pseudorapidity. Then, for each event, we determined the region where the delayed photon would enter the ECAL. With this we calculated the absolute time $t'$ using the heavy neutrino and photon momenta, and the heavy neutrino decay position $\vec r$. Finally, one has $t_\gamma=t'-t_0$. We determined that the $\alpha_{NB}/\Lambda$ values shown on the third row of Table~\ref{Table:Benchmarks} maximized the number of contained events featuring a large $|\Delta z_\gamma|$ and $t_\gamma$.

\begin{figure}[tp] 
\centering
\includegraphics[width=0.45\textwidth]{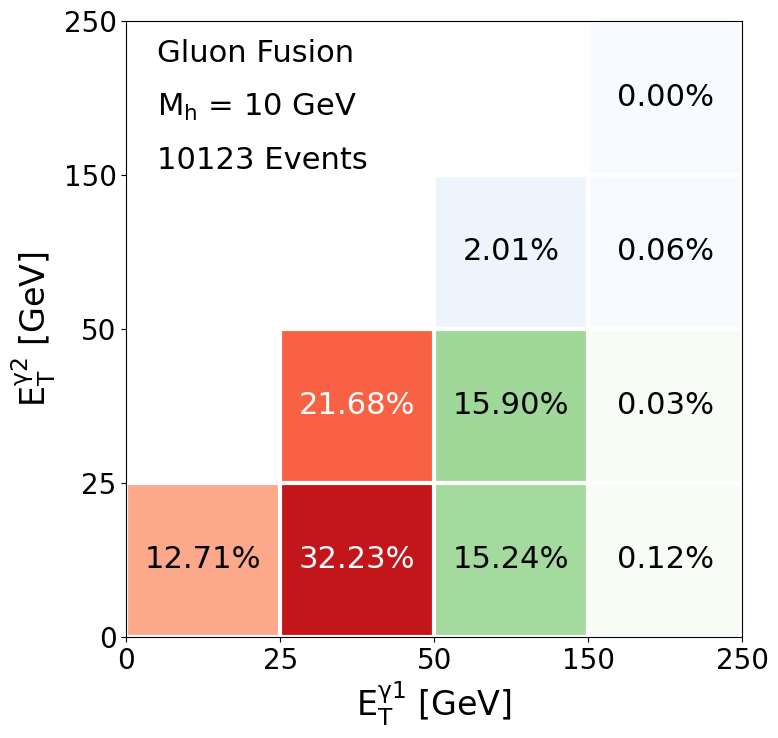} \hfill
\includegraphics[width=0.45\textwidth]{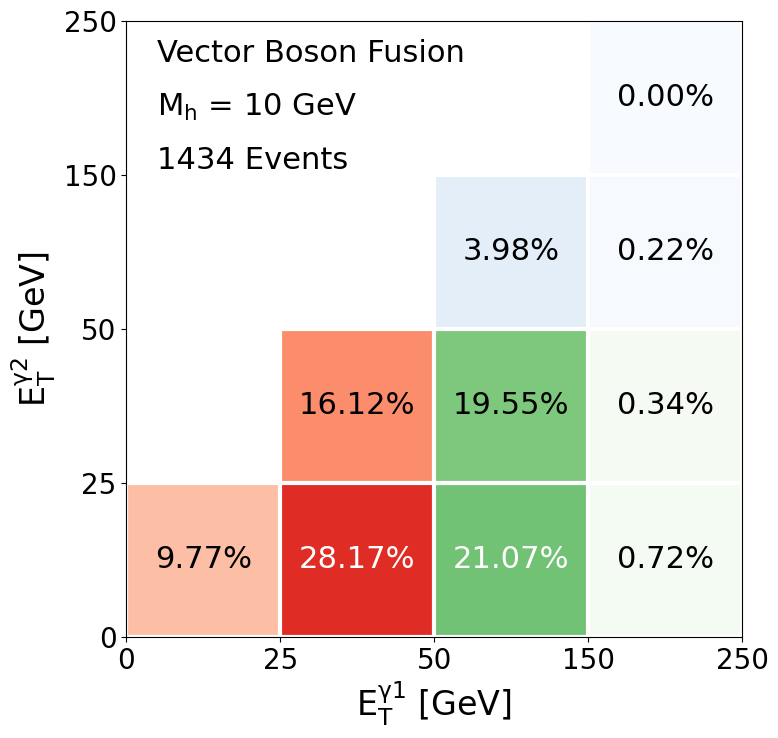} \vspace{3mm} \\
\includegraphics[width=0.45\textwidth]{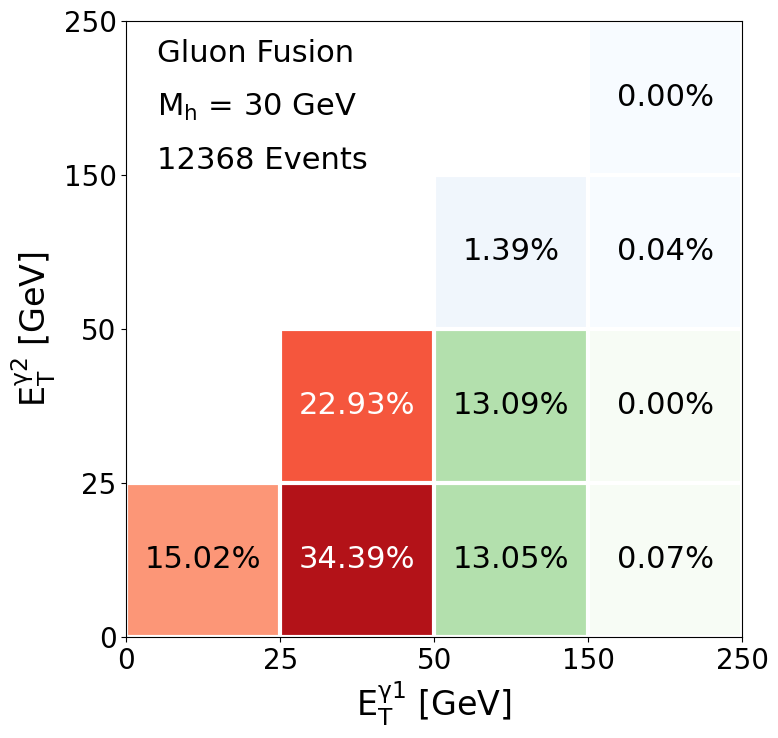} \hfill
\includegraphics[width=0.45\textwidth]{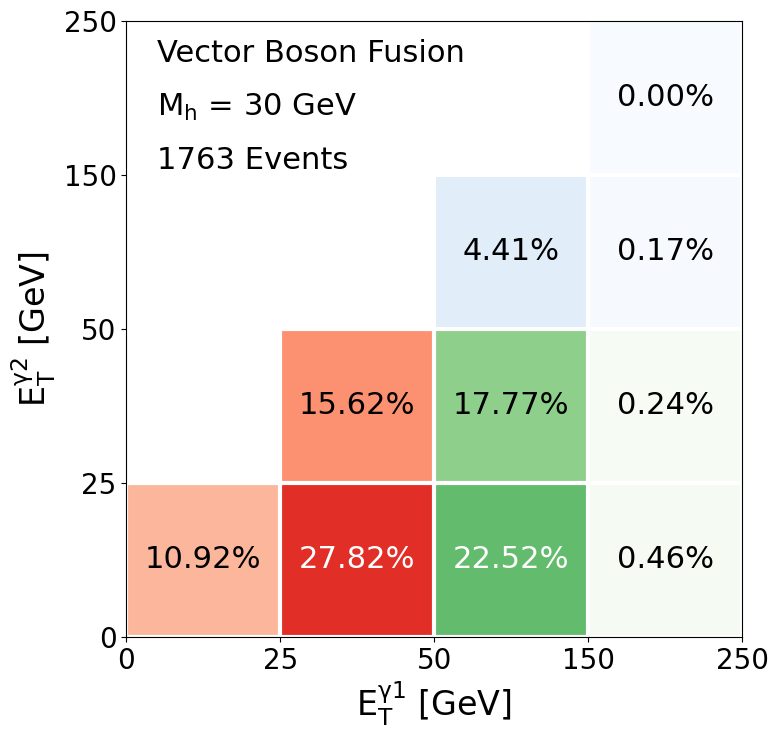} \vspace{3mm} \\
\includegraphics[width=0.45\textwidth]{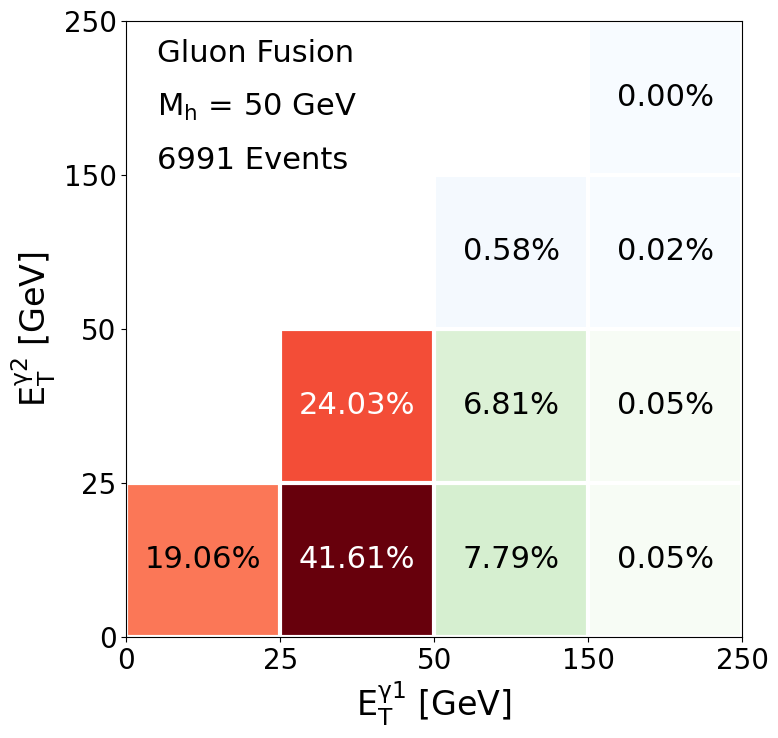} \hfill
\includegraphics[width=0.45\textwidth]{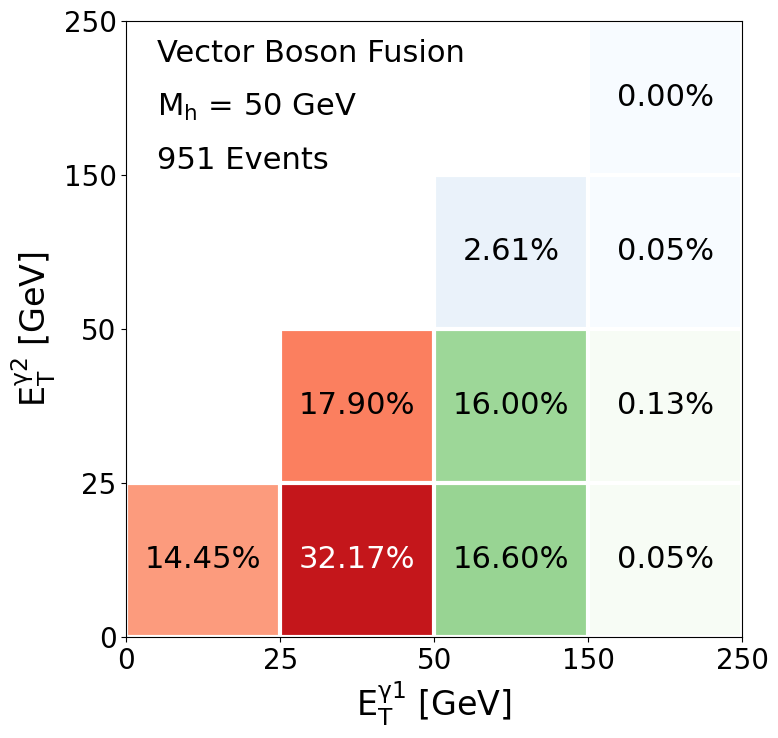}
\caption{Percentage of two-photon events coming from heavy neutrino decay, for different energy bins. Higgs production via GF (VBF) is shown on the left (right) column. Results for $M_h=10,\,30,\,50$~GeV are shown on the top, center, bottom rows, respectively. Number of events refers to those where both photons are contained within the detector. See text for details.} 
\label{fig:HeatMaps}
\end{figure}
In the following, we will argue that this search is not particularly sensitive to our model. The first problem we find are the very large 50 GeV cuts on photon energy. In particular, in Figure~\ref{fig:HeatMaps} we define energy bins for the two photons, such that one can build ``heat maps" indicating what is the most likely distribution of the di-photon energy of our signal events. In the Figure, the red bins are excluded by cuts on both photons, the green bins are excluded by cuts on the sub-leading photon, while the blue bins are allowed. For GF (VBF), between $67\%$ and $85\%$ ($54\%$ and $65\%$) of our signal events do not pass any of the cuts on the two photons, and only between $0.6\%$ and $2.1\%$ ($2.7\%$ and $4.6\%$) of them pass both. Considering the number of contained events, as reported on each panel, we expect a total of 210, 177, and 42 (60, 81 and 25) events passing both cuts, for $M_h=10,\,30,\,50$~GeV respectively, assuming GF (VBF) Higgs production.

\begin{figure}[tp]
\centering
\includegraphics[width=0.48\textwidth]{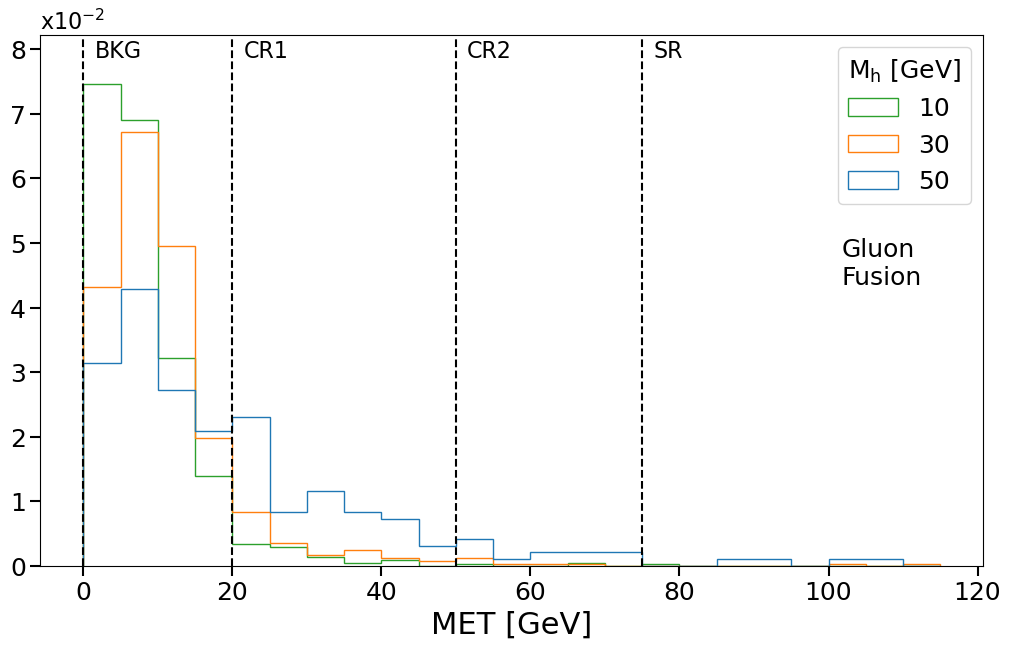} \hfill
\includegraphics[width=0.48\textwidth]{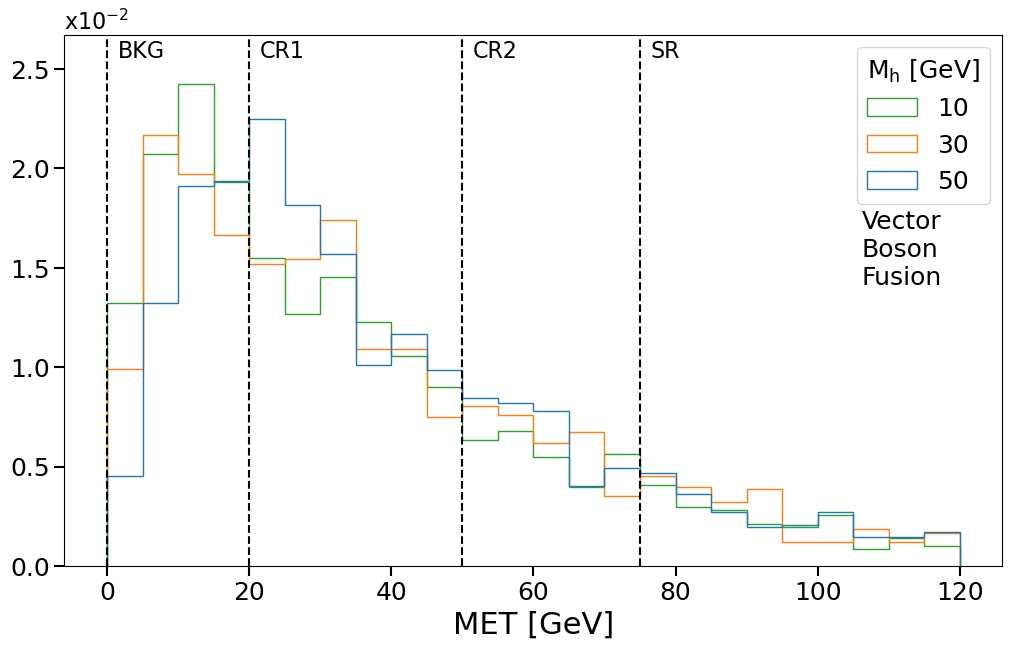}
\caption{Missing energy distribution of heavy neutrino decays, for several $M_h$. Higgs production via gluon fusion (vector boson fusion) is shown on the left (right). } 
\label{fig:MET8}
\end{figure}
Even though the number of events passing the photon energy cuts is not negligible, we find further problems with the definition of background, control and signal regions. To illustrate this, Figure~\ref{fig:MET8} shows the MET distribution of the different generated signal events samples, after applying all cuts indicated above. The vertical lines indicate the MET regions defined in the ATLAS analysis, as shown in Table~\ref{Table:Regions1}. It can be clearly seen that the GF production channel (left panel) is not useful at all, with practically no events categorized in the SR region. According to our results, for 10 and 30~GeV masses we would have no signal, and the $M_h=50$~GeV scenario would have only one event. Considering that we are not including detector effects at this point, it is clear that that this search would not be able to probe this model if only GF was considered.

The right panel of Figure~\ref{fig:MET8} shows that the VBF channel is somewhat more promising. Even though we have less events being generated, due to the lower cross-section, we find a much larger efficiency. According to our results, we would have 8, 13, and 4 events in the SR region, for $M_h=10,\,30,\,50$~GeV respectively. However, this must be compared with the 386 events in the SR region reported by~\cite{ATLAS:2014kbb}. Morover, we find that, within the six $|\Delta z_\gamma|$ categories defined in the analysis, all of our events lie on bins dominated by the background. In particular, no signal region events lie in bins of $t_\gamma$ larger than $\sim1$~ns. Taking all of these issues into consideration, we find it unlikely that our few surviving events will be of any use to place bounds on the model. It must be emphasized that the MET distribution is not improved even if the cuts on the photon energy are relaxed.

\begin{figure}[tp]
\centering
\includegraphics[width=0.48\textwidth]{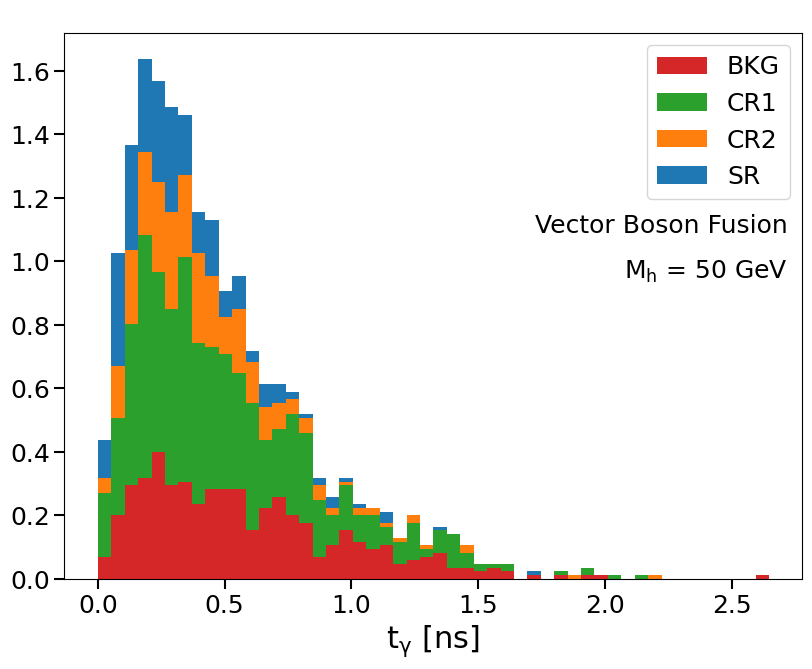} \hfill
\includegraphics[width=0.48\textwidth]{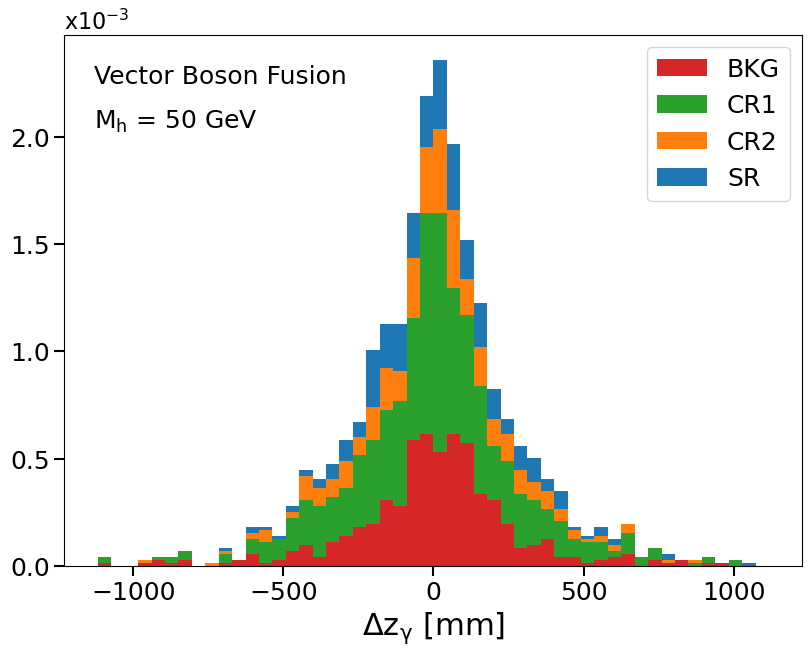}
\caption{Distributions of $t_\gamma$ (left) and $\Delta z_\gamma$ for the $M_h=50$~GeV case, considering VBF production. We show in red, green, orange and blue the generated signal events classified in the BKG, CR1, CR2 and SR regions, respectively.} 
\label{fig:8TeVdistribs}
\end{figure}

Finally, we show in Figure~\ref{fig:8TeVdistribs} the $t_\gamma$ and $\Delta z_\gamma$ distributions for our VBF generated signal events, for $M_h=50$~GeV, after passing all cuts. As above, depending on the MET, these are classified into BKG, CR or SR regions. On the left panel, we see that most events have $t_\gamma\lesssim0.5$~ns. This makes sense, since in order to pass the 50~GeV cuts, the $N_h$ need to be very boosted, leading to non-delayed photons. In addition, the vast majority of our signal events with $t_\gamma>1$~ns are classified into BKG or CR regions. This gives further evidence that the region definition is not optimal for our model. On the right panel, even though most events lie around zero, one finds $|\Delta z_\gamma|$ values up to 500~mm. Such large values can be found in all regions, with most of them falling in CR1. Scenarios with lighter $M_h$, or using GF production, have lesser values of both $t_\gamma$ and $|\Delta z_\gamma|$.

We conclude that even if a photon pair from long-lived $N_h$ passed the stringent energy cuts, and even if they also had large $t_\gamma$ and/or $|\Delta z_\gamma|$, it is more likely that such an event would be assigned to the background or control region, than to the signal region. Thus, this is strongly suggesting this search is not optimal for studying our model.

\subsection{13 TeV ATLAS Search (2021)}
\label{sec:Thesis}

Even though there are no published papers featuring non-pointing photons at energies above 8 TeV, there does exist a somewhat recent thesis addressing this kind of search, for 13 TeV~\cite{Mahon:2021bai}\footnote{Shortly before this work was completed, the ATLAS collaboration released a conference note related to this analysis~\cite{ATLAS:2022bsa}.}. This work presents a search for exotic decays of the Higgs boson, targeting the open phase space of its decays to some intermediate invisible long-lived particle with displaced photons in the final state. The search uses 139~fb$^{-1}$ of data in $13$ TeV collisions from ATLAS, collected between 2015 and 2018, probably representing the first displaced photon analysis performed using the full LHC Run 2 dataset, with seven times more integrated luminosity compared to the previous ATLAS result \cite{ATLAS:2014kbb}. The search is designed to have sensitivity to softer non-pointing photons signatures than in the previous study, for the same GMSB signal, but with neutralinos from Higgs decays. To this end, they trigger the measurement with an associated lepton from Higgs production, allowing for a relaxation of the photon energy cuts used before. Thus, the final state contains at least one electron or muon, at least one photon, and MET. 

Again, photons are identified with the loose photon ID algorithm, this time selected if they have $p_T > 10$ GeV and $|\eta|< 2.37$ (excluding the transition region). These are considered isolated if all ECAL topo-clusters within a fixed $\Delta R = 0.2$ cone (excluding the core) are less than $6.5\%$ of the total cluster $p_T$. In addition, track isolation requires that the fixed-radius $R = 0.2$ cone, excluding the photon candidate itself, must contain less than $5\%$ of the total object $p_T$. For the photon timing, only the cell in the ECAL middle layer with the maximum energy deposit is used ($E_{\rm cell}$) and a cut is placed at either 7 or 10 GeV, depending on the GMSB signal point being analyzed, as the low-energy photons degrade the analysis sensitivity. With this, photons likely originating from out-of-time pileup are avoided by discarding events with $t_{\gamma}>12$~ns.

Another important difference in this new analysis is that events are categorized into either one-photon or multi-photon channels. In both, all photons must satisfy the aforementioned selection requirements. Moreover, it is the leading barrel photon which is used for the fit, instead of that with largest relative arrival time $t_\gamma$.

This search also requires identifying the leptons coming from Higgs production. Their selection criteria, including isolation and geometric cuts can be found in \cite{Mahon:2021bai}.

Depending on the MET, the selected sample is divided into a Control, Validation, and Signal regions, which we denote BKG, CR and SR respectively, similar to~\cite{ATLAS:2014kbb} (see Table~\ref{Table:Regions1}). Here, one can see that the SR region allows much smaller values of MET. Finally, the sample passing the selection cuts is again divided into $|\Delta z_{\gamma}|$ categories, with each divided into $t_{\gamma}$ bins. These are chosen to optimize the sensitivity to the same GMSB signal.

As one can see, this thesis satisfies our wishlist from the 8 TeV search, namely, lesser cuts on the photon energy and a signal region allowing for smaller MET. In the following, we will recast the cuts described above, and apply them on our model events. However, in order to compare both 8 and 13 TeV searches, we generate Higgs events in the VBF channel, instead of producing Higgs and leptons. The measurement is triggered by the VBF jets, imposing the same cuts as in~\cite{Jones-Perez:2019plk}.

Our procedure has two stages. First, we take the \texttt{HepMC} data, and apply the VBF, photon $p_T$, and photon $\eta$ cuts described above. We only consider photon pairs produced both within the inner detector. To be conservative, for a given photon the isolation cuts require no leptons or other photons with $p_T>10$~GeV, and no jets with $p_T>20$~GeV, within a $\Delta R=0.2$ cone. 

With this sample, both $|\Delta z_\gamma|$ and $t_\gamma$ are calculated for the most energetic photon in the barrel. The calculation is performed in the same way as for the 8 TeV search, this time including experimental uncertainties.  The uncertainty in $|\Delta z_\gamma|$ is replicated by applying a gaussian smear, using the resolution shown in Figure 1 of~\cite{ATLAS:2014kbb}. For the $t_\gamma$ uncertainty, which depends on $E_{\rm cell}$, we define the latter as $30\%$ of the true photon energy\footnote{Generally around 15–40 \% of the energy in the entire EM cluster is deposited in the middle layer cell with the maximum energy \cite{Mahon:2021bai}.}. With this, the uncertainty is implemented again via a gaussian smear, with the resolution taken from both Eq.~(5.5) and Table~A.2 in~\cite{Mahon:2021bai}.

For the second stage of the procedure, the events are separated into single photon and multi-photon sets, and each set is further categorized into the $|\Delta z_\gamma|$ and $t_\gamma$ bins. To appropriately simulate detector effects, the events in each bin are separately processed by \texttt{Delphes}. For consistency, the VBF and photon cuts are re-applied. However, the photon isolation was implemented by modifying the appropriate module of the \texttt{Delphes} ATLAS card, setting $\Delta R=0.2$ and the track isolation \texttt{PTRatioMax} to $0.05$. In addition, the cut on $E_{\rm cell}>10$~GeV was applied, this time using $30\%$ of the reconstructed photon energy. Finally, the number of events on each bin was re-adjusted, in case the number of photons reported by \texttt{Delphes} was diferent than that obtained in the first stage.

\begin{figure}[tp]
\centering
\includegraphics[width=0.48\textwidth]{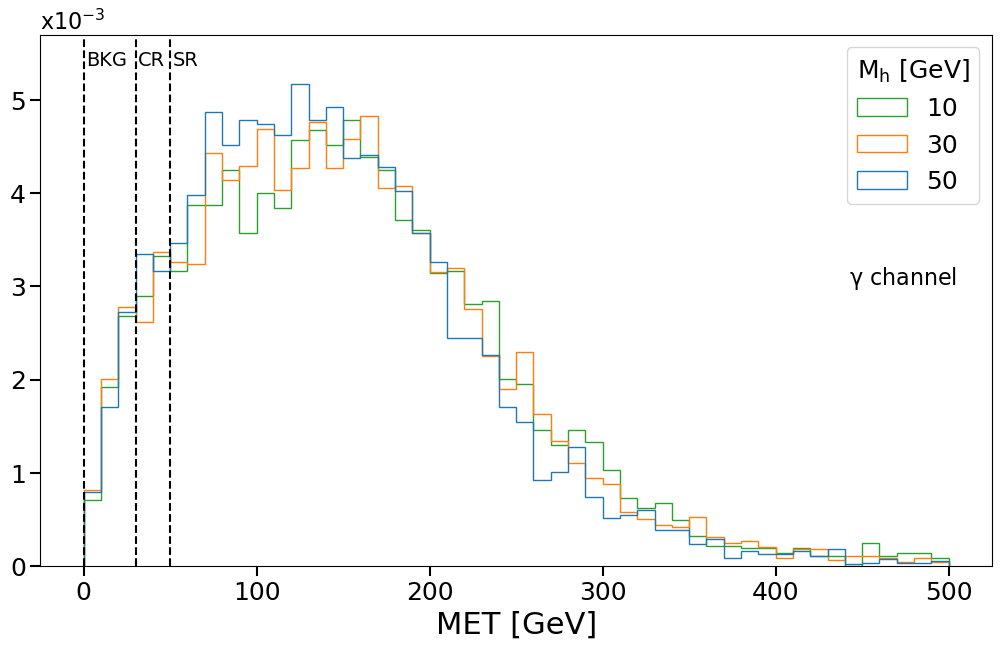} \hfill
\includegraphics[width=0.48\textwidth]{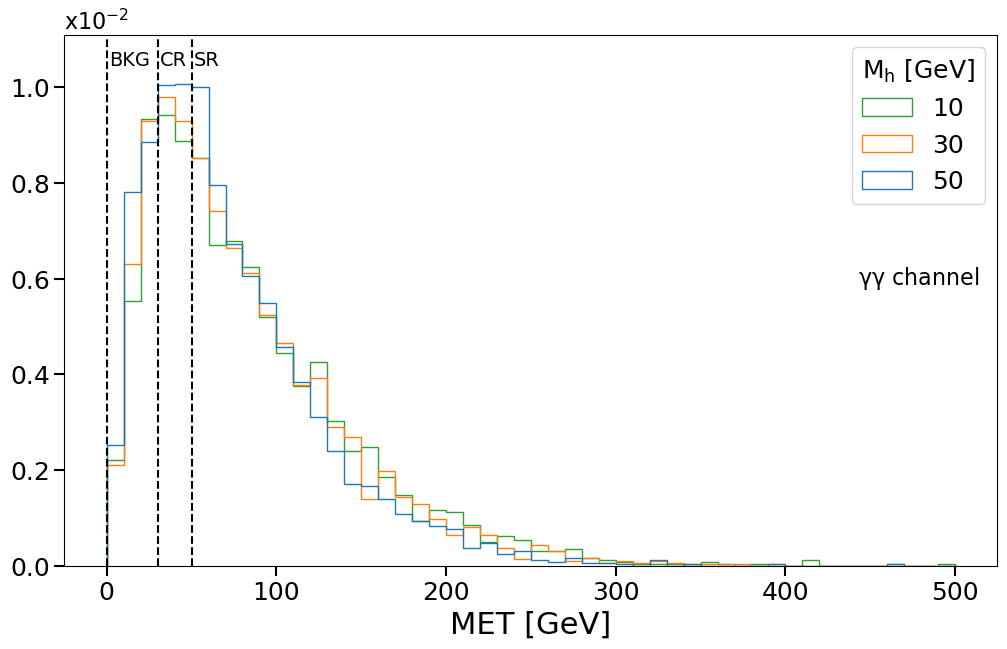}
\caption{Missing energy distribution of heavy neutrino decays in 13 TeV analysis, for several $M_h$.  Single (multi-) photon set is shown on the left (right). } 
\label{fig:MET13}
\end{figure}
We show in Figure~\ref{fig:MET13} the MET distribution for both single and multi-photon sets. By comparing with Figure~\ref{fig:MET8}, we immediately confirm that the cuts proposed in this search are much more appropriate for our signal. In particular, the peak of the MET distribution in the single photon set is very far from the BKG and CR regions. We find 207, 241, and 215 events in the SR region, for $M_h=10,\,30,\,50$~GeV, respectively. In contrast, the multi-photon set has much more events contaminating the BKG and CR regions. For this set, the SR has 104, 112 and 92 events, for $M_h=10,\,30,\,50$~GeV, respectively. In any case, both channels imply a significant improvement with respect to the 8 TeV search. Moreover, these numbers are comparable to those shown in Table~7.5 of~\cite{Mahon:2021bai}, suggesting that this analysis could effectively be used to probe our model.

\begin{figure}[p]
\centering
\includegraphics[width=0.32\textwidth]{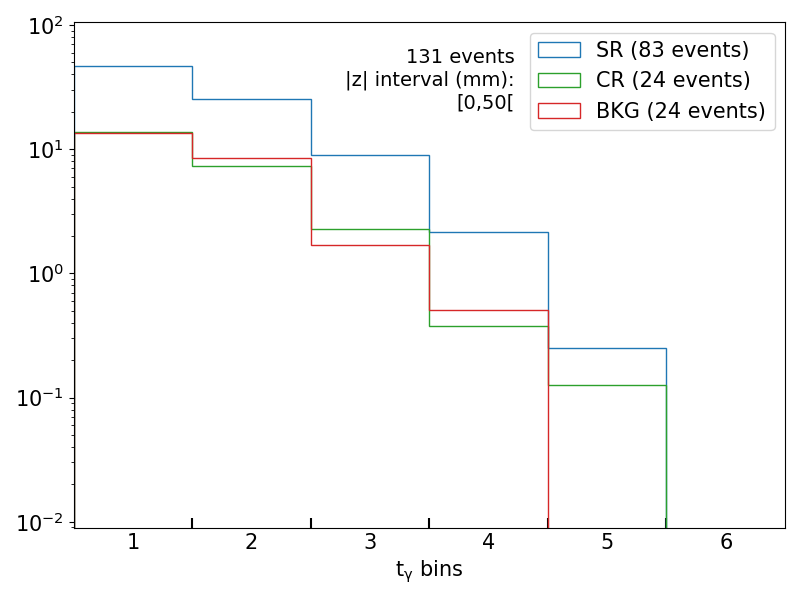} \hfill
\includegraphics[width=0.32\textwidth]{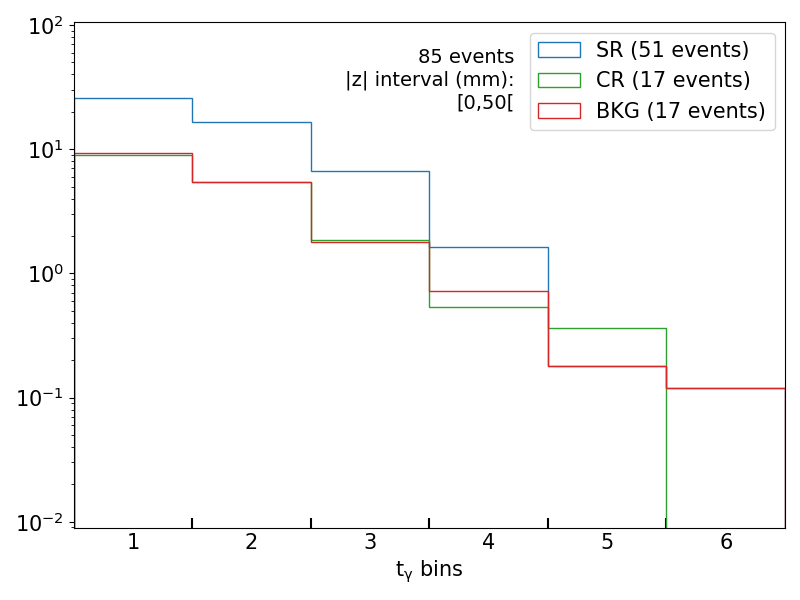} \hfill
\includegraphics[width=0.32\textwidth]{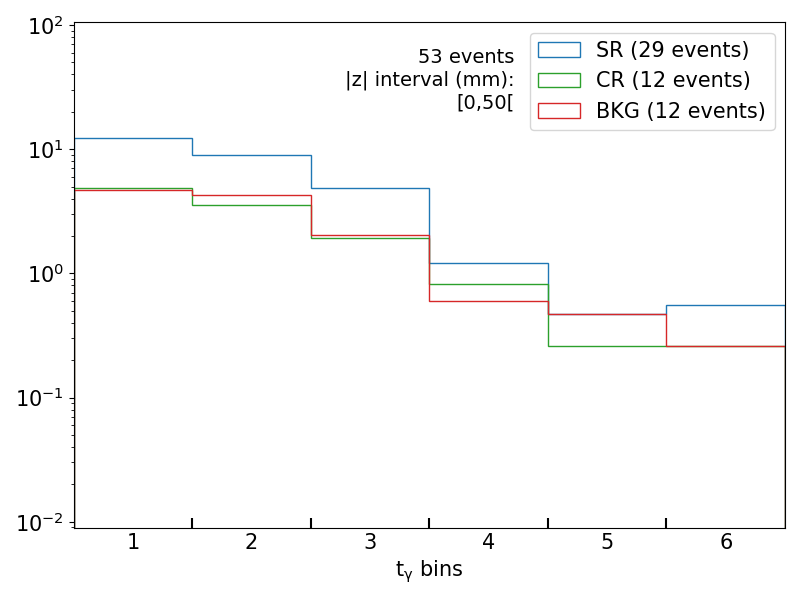} \\
\includegraphics[width=0.33\textwidth]{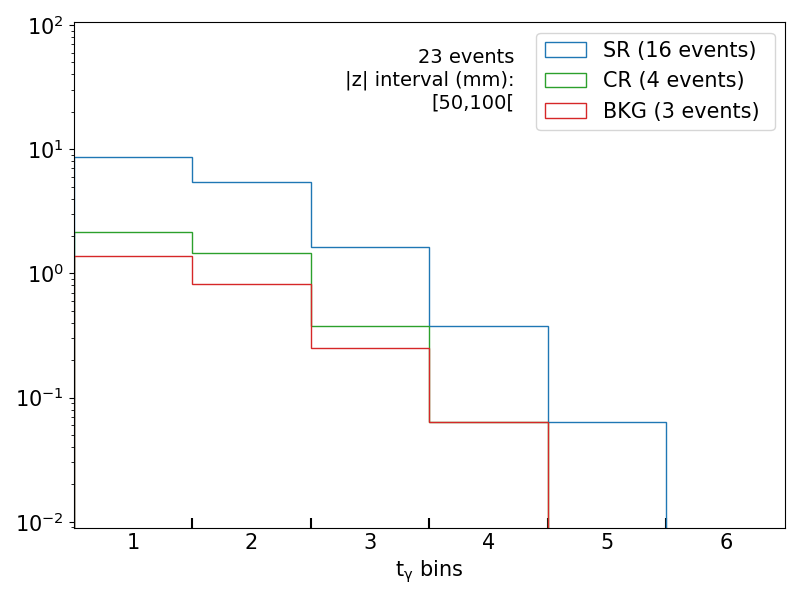} \hfill
\includegraphics[width=0.32\textwidth]{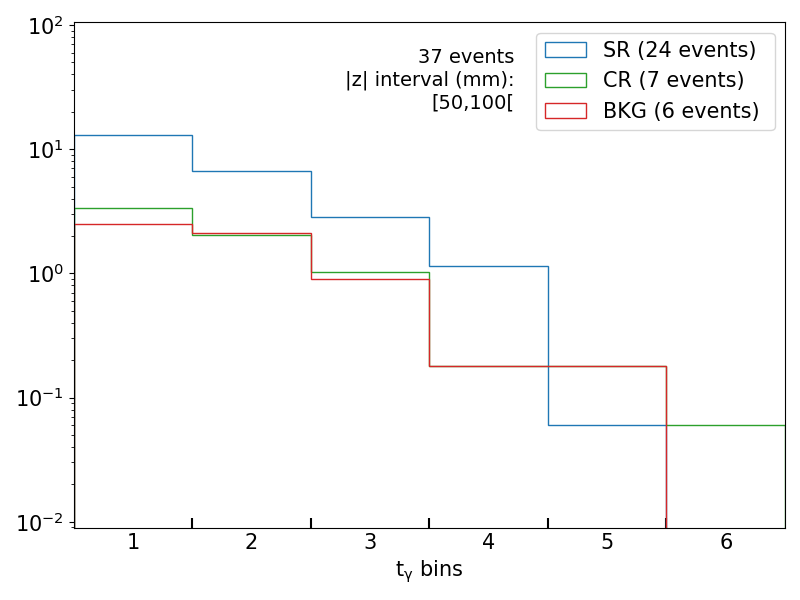} \hfill
\includegraphics[width=0.32\textwidth]{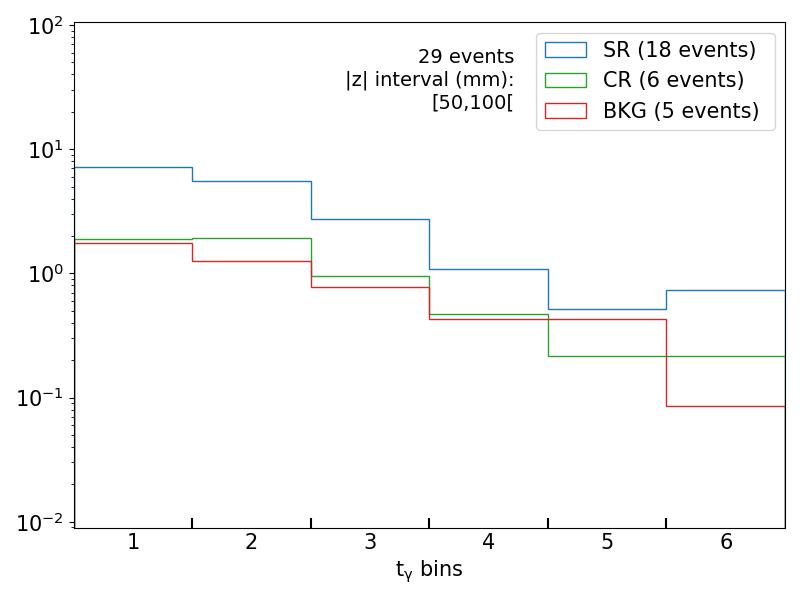} \\
\includegraphics[width=0.32\textwidth]{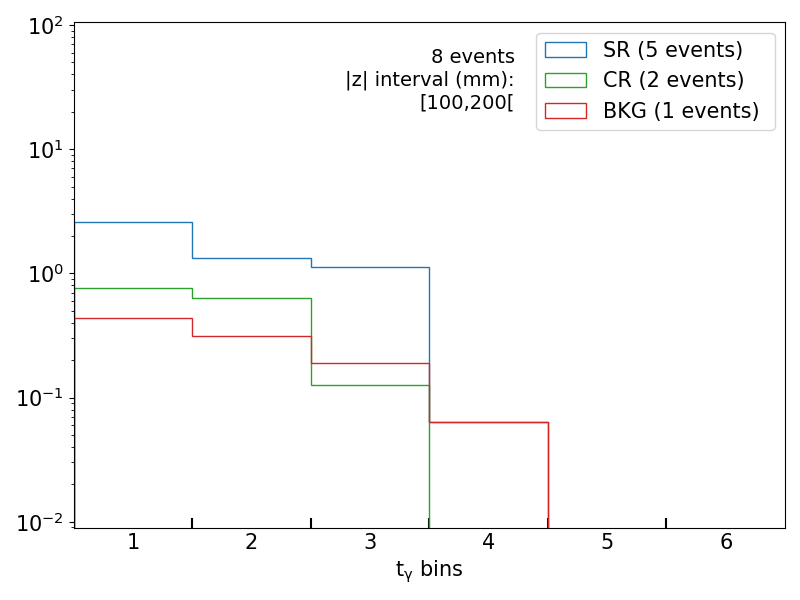} \hfill
\includegraphics[width=0.32\textwidth]{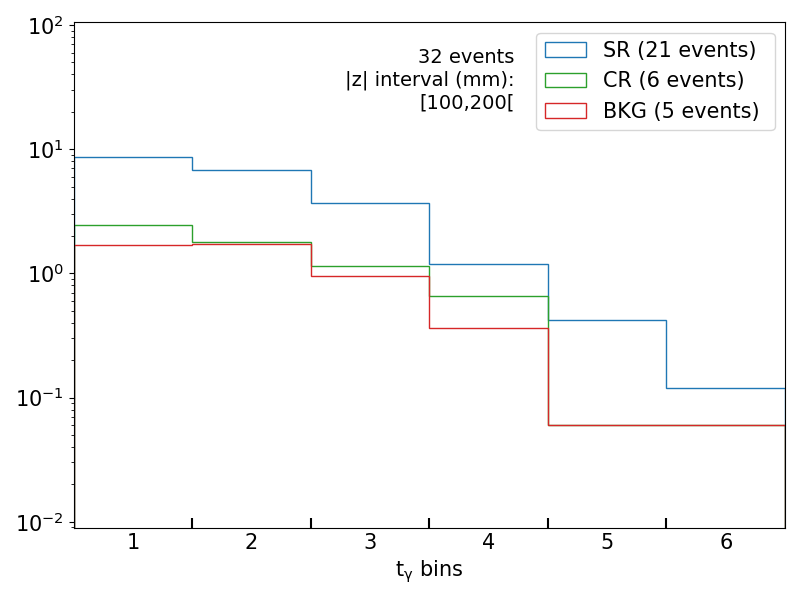} \hfill
\includegraphics[width=0.32\textwidth]{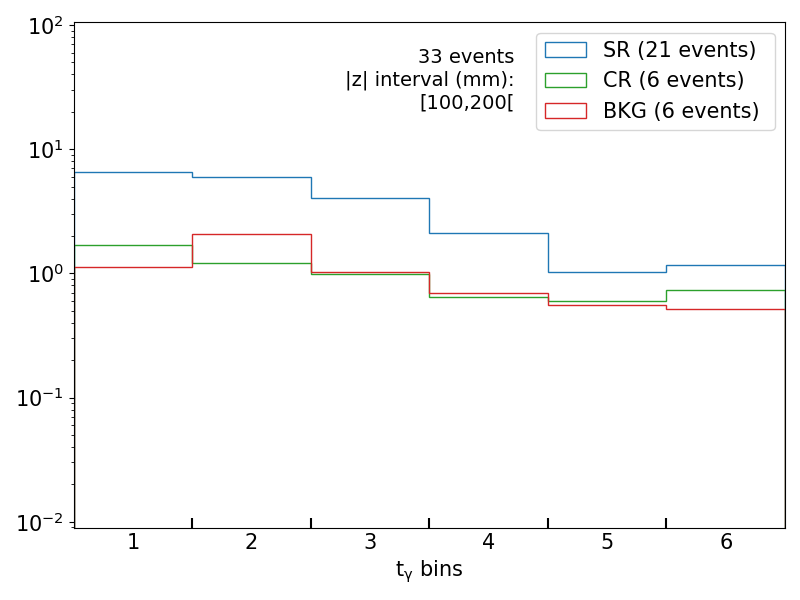} \\
\includegraphics[width=0.32\textwidth]{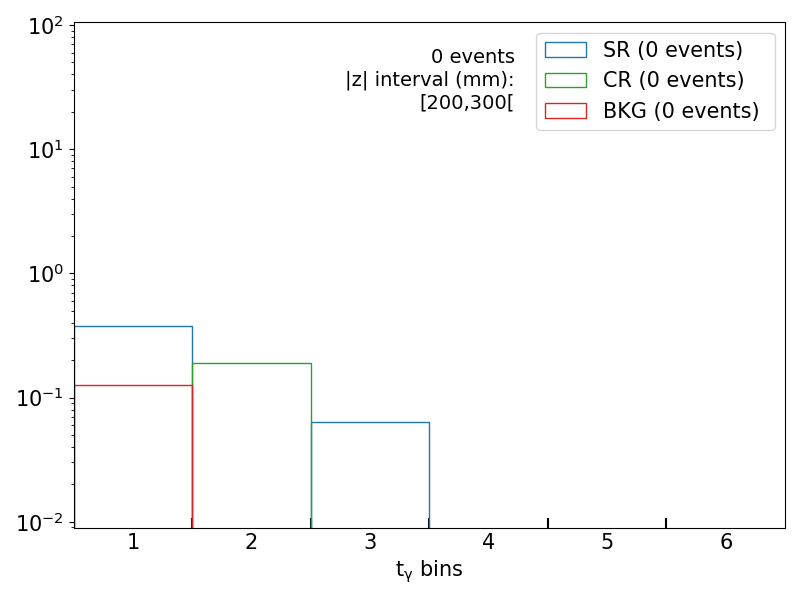} \hfill
\includegraphics[width=0.32\textwidth]{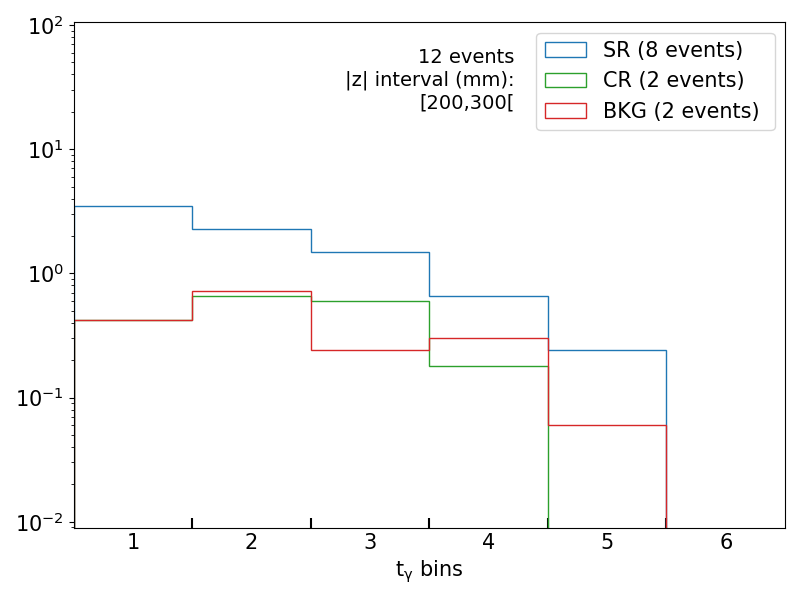} \hfill
\includegraphics[width=0.32\textwidth]{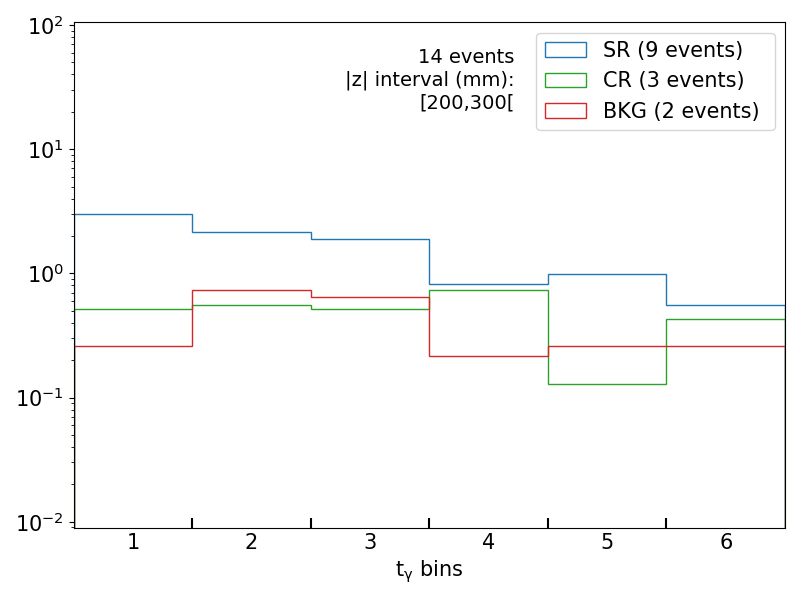} \\
\includegraphics[width=0.32\textwidth]{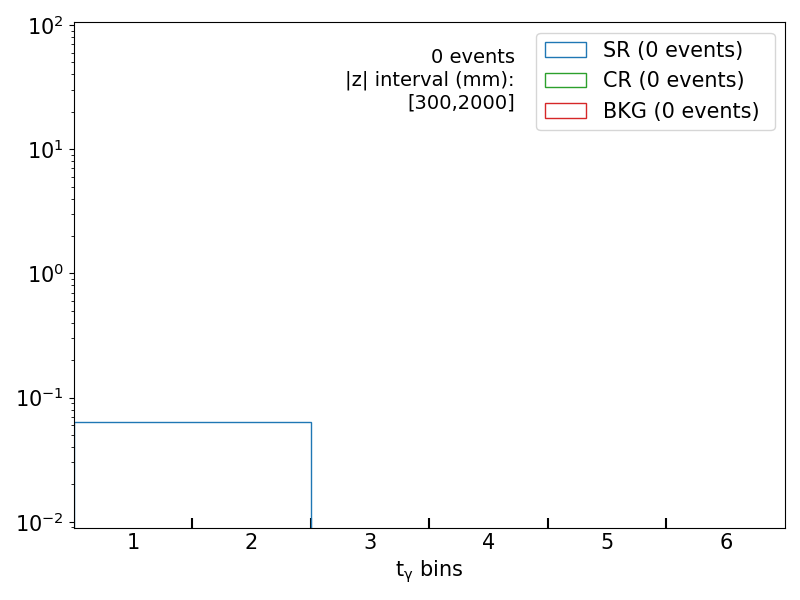} \hfill
\includegraphics[width=0.32\textwidth]{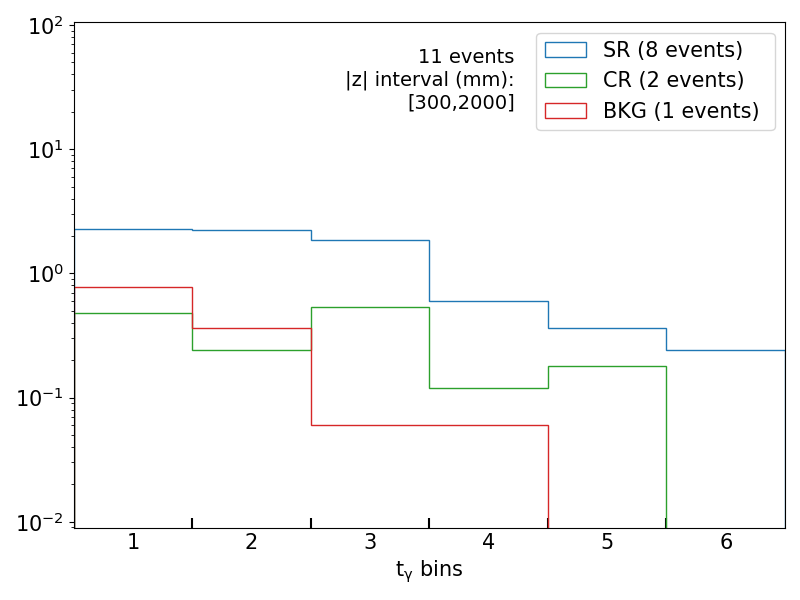} \hfill
\includegraphics[width=0.32\textwidth]{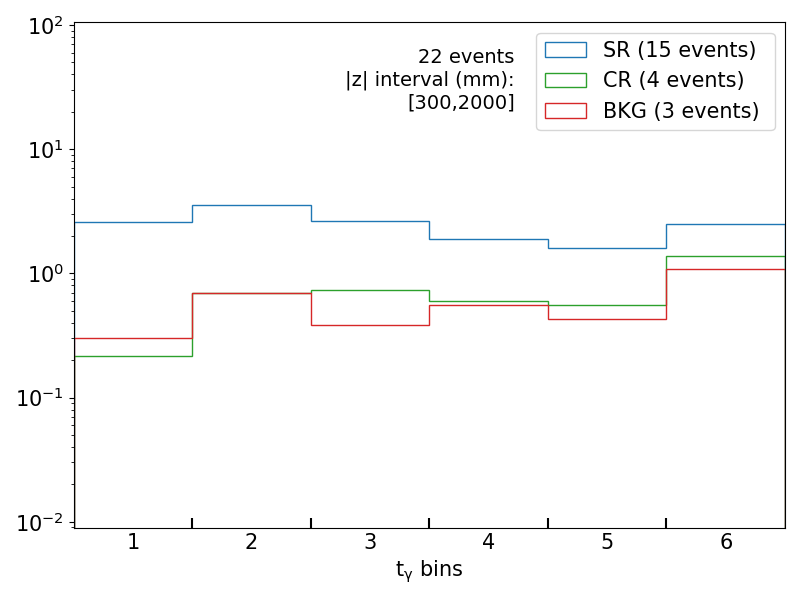}
\caption{Event distribution for the multi-photon category, for $M_h=10,\,30,\,50$~GeV on the left, center and right columns, respectively. Each row indicates a different $|\Delta z_\gamma|$ range. The six bins indicate $t_\gamma$ values between 0, 0.2, 0.4, 0.6, 0.8, 1 and 12~ns. Events in different regions are not stacked.} 
\label{fig:13bins2}
\end{figure}
Finally, we show in Figure~\ref{fig:13bins2} the event distribution for the multi-photon set. The plot follows the binning in $|\Delta z_\gamma|$ and $t_\gamma$ described above. In all plots we show the events in the BKG, CR and SR regions. From here, we can see that even though the peak of the multi-photon distribution is shared by all regions, as shown in Figure~\ref{fig:MET13}, we still have a significant number of events in the SR region, which can also have large $t_\gamma$ and $|\Delta z_\gamma|$ values.

Let us turn to the large non-pointing $|\Delta z_\gamma|$ categories. Here, the photon momentum must have a relatively large angle with respect to the $N_h$ momentum. Thus, if the heavy neutrino is highly boosted, it is expected that $|\Delta z_\gamma|$ will be small, as the photons will be emitted following the $N_h$ boost direction. In this sense, lighter $N_h$ samples are predicted to have less events at large $|\Delta z_\gamma|$ bins. Figure~\ref{fig:13bins2} confirms our expectations, for example, when $M_h=10$~GeV the large $|\Delta z_\gamma|$ categories have virtually no events. Heavier $N_h$ are produced with a much smaller boost, such that it is more likely to find events with large $|\Delta z_\gamma|$.

In order to obtain a large photon $t_\gamma$, one needs a slow-moving heavy neutrino with the daughter photon momentum again having a large angle with respect to that of the $N_h$. As before, this is easier to achieve for the heaviest $N_h$ in our analysis. Figure~\ref{fig:13bins2} shows that, for $M_h=50$~GeV, one can find events with large $t_\gamma$ in all $|\Delta z_\gamma|$ categories.

If one naively compares Figure~\ref{fig:13bins2} with Figure~10.1 of~\cite{Mahon:2021bai}, it would still seem very difficult to probe our model. In fact, the background is very large, specially in the small $t_\gamma$ and $|\Delta z_\gamma|$ bins. However, it is necessary to be reminded that our analysis uses a VBF trigger, instead of the associated lepton trigger in~\cite{Mahon:2021bai}. As shown in many works~\cite{Dutta:2012xe, Delannoy:2013ata,Liu:2014lda,Bambhaniya:2015wna,Andres:2016xbe,Andres:2017daw,Dutta:2017lny,Florez:2018ojp,Florez:2019tqr,Duque-Escobar:2021rdz}, the VBF topology is very convenient for background reduction. 


To quantify the sensitivity to our model, we now refer to a recent conference note following this analysis~\cite{ATLAS:2022bsa}, released by ATLAS shortly before this work was completed. Apart from the aforementioned analysis, the note includes a model-independent prediction of the background for the bins with largest $|\Delta z_\gamma|$ and $t_\gamma$ values (see Table~II of~\cite{ATLAS:2022bsa}), which can be used to estimate the sensitivity. For the corresponding bin of the multi-photon channel, the $M_h=50$~GeV scenario predicts 2.5 signal events. Thus, if we take this number and compare directly with the estimated background~\cite{Cowan:2010js}, we would achieve a local significance $Z_0=2.8$ standard deviations. This result is already encouraging. Actually, if the background reduction from the VBF trigger was, for instance, of one order of magnitude, the local significance would rise to $Z_0=4.2$.




We finally note that the single photon set has a similar distribution, but about twice as many events. However, backgrounds are also expected to be larger by at least one order of magnitude, so we expect the multi-photon set to be more useful when probing our model.

To summarize, we consider our findings to be very promising, and recommend the experimental community to take into account the $H\to N_h\, N_h$ production mode via VBF in the future. We leave the analysis of $N_h$ production with the original associated lepton trigger used in~\cite{ATLAS:2022bsa}, and with this a rigorous exclusion of $\alpha_{N\phi}$ values, for our next project.

\section{Conclusions}

In this work we have revisited the Type-I Seesaw enlarged with dimension-5 effective operators. The model is characterized by new couplings between the sterile neutrino states and SM neutral bosons, via the Anisimov-Graesser and dipole operators. These lead to new interactions involving light neutrinos via mixing.

After presenting our setup, we have calculated the decay width of the heavy neutrinos. We find that, although usually dominated by $N_h\to\nu\,\gamma$, the standard Seesaw three-body decays have a very important role when the heavy neutrino masses are above the GeV, and the dipole coupling $\alpha_{NB}/\Lambda$ is not too large. We have also calculated the modification to the three-body partial widths from the dipole coupling, and find that the impact on the decay length does not exceed $10\%$.

We then review existing constraints on the dipole coupling, and re-evaluate the LEP bounds from $e^+e^-\to N_h\,\nu_\ell$ production. We find that if the sterile-light mixing is not enhanced from its naive Seesaw value, there are no bounds on the coupling. However, if the mixing is large the bound could rule out at most $\alpha_{NB}/\Lambda\gtrsim10^{-8}$~GeV$^{-1}$, for $M_h\sim1$~GeV.

Finally, we turn to non-pointing photon searches at the LHC. These searches could probe regimes where an exotic Higgs decay pair produces long-lived heavy neutrinos, disintegrating later into a photon and a light neutrino. The latest published paper on this topic corresponds to an 8 TeV ATLAS search (2014). We also found a 13 TeV ATLAS search in the form of a PhD thesis (2021), with an associated conference note (2022). We recasted both, concluding that the 8 TeV search is not sensitive to our model. Our implementation of the 13 TeV analysis differed from the original search in the sense that we chose a VBF trigger, instead of single lepton for Higgs production. We found a competitive number of events passing all cuts, suggesting that this kind of analysis could place future bounds on the Higgs branching ratio to these long-lived neutrino states.

\acknowledgments
We would like to thank Claude Duhr and Olivier Mattelaer for assistance when debugging our \texttt{FeynRules} model. We acknowledge the financial support of Pro CIENCIA, CONCYTEC (Peru's National Council for Science and Technology), Contract 123-2020-FONDECYT and PUCP DARI Groups Research Fund 2019. F.D.\ and J.J.P.\ acknowledge funding by the {\it Direcci\'on de Gesti\'on de la Investigaci\'on} at PUCP, through grants No.\ DGI-2019-3-0044 and DGI-2021-C-0020.  L.D. is funded by PEDECIBA-F\'isica. 

\appendix

\section{Numerical Tools}
\label{app:NumericalTools}

The partial widths in Appendix~\ref{app:partial_widths}, as well as the $N_h\,\nu_\ell$ production cross-section at LEP, were calculated with the assistance of \texttt{FeynCalc 9.3.1}~\cite{Mertig:1990an,Shtabovenko:2016sxi,Shtabovenko:2020gxv}. For the widths, our numerical results were checked with \texttt{MadWidth}~\cite{Alwall:2014bza}. The $\Delta_{\rm QCD}$ function was evaluated with \texttt{RunDec 3.1}~\cite{Chetyrkin:2000yt,Herren:2017osy}. The cross-section was integrated using the \texttt{Vegas} subroutine within the \texttt{Cuba 4.2} library~\cite{Hahn:2004fe}.

For our LHC calculations, we modified the \texttt{HeavyN} model~\cite{Alva:2014gxa,Degrande:2016aje} in \texttt{FeynRules 2.3.43}~\cite{Christensen:2008py,Alloul:2013bka}, adding the effective operators. Events were generated using \texttt{MadGraph5\_aMC@NLO 2.9.7}~\cite{Alwall:2014hca}, which uses \texttt{LHAPDF6}~\cite{Buckley:2014ana}, followed by \texttt{PYTHIA 8.244}~\cite{Sjostrand:2006za}, which carries out the showering and hadronization. For the detector simulation, we used \texttt{Delphes 3.5.0}~\cite{deFavereau:2013fsa}, which depends on \texttt{FastJet 3.4.0}~\cite{Cacciari:2011ma}.

\section{Heavy neutrino three-body partial widths}
\label{app:partial_widths}

This Appendix collects the full fomulae for heavy neutrino three-body decays in the GeV range. We remind the reader that $\nu_a$ refers to active interaction eigenstates, while $\nu_\ell$ refers to light mass eigenstates. In what follows we report only the partial widths modified by the effective operator, that is, those where either the $Z$ or photon participate. For purely $W$-mediated decays like $N_h\to l_a\,q\,q'$ and $N_h\to l_a\,l_{a'}\nu_{a'}$, we refer the reader to the appropriate literature~\cite{Atre:2009rg,Kovalenko:2009td,Bondarenko:2018ptm,Coloma:2020lgy}. 

For three-body decays, we define:
\begin{align}
g^u_L =& \tfrac{1}{2}-\tfrac{2}{3}s_W^2&
g^u_R =& -\tfrac{2}{3}s_W^2\\
g^d_L =& -\tfrac{1}{2}+\tfrac{1}{3}s_W^2&
g^d_R =& \tfrac{1}{3}s_W^2 \\
g^l_L =& -\tfrac{1}{2}+s_W^2&
g^l_R =& s^2_W
\end{align}

We also use the following functions:
\begin{equation}
\lambda(a,\,b,\,c) = a^2+b^2+c^2-2ab-2ac-2bc   
\end{equation}

\begingroup
\allowdisplaybreaks
\begin{align}
\Gamma_{1}^{ji}(a,b,c)&=12\int_{(b+c)^{2}}^{(1-a)^{2}} (1+a^{2}-s)^{j}\,(s-b^2-c^2)\,\lambda^{1/2}(1,a^{2},s)\,\lambda^{1/2}(s,b^{2},c^{2})\frac{\diff s}{s^{1+i/2}}\\
\nonumber \\
\Gamma_{5}^{ji}(a,b,c)&=24\,b\,c\int_{(b+c)^{2}}^{(1-a)^{2}} (1+a^{2}-s)^{j}\,\lambda^{1/2}(1,a^{2},s)\,\lambda^{1/2}(s,b^{2},c^{2})\frac{\diff s}{s^{1+i/2}} \\
\nonumber \\
\Gamma_{2}^{ji}(a,b,c)&=4\int_{(b+c)^{2}}^{(1-a)^{2}} (1+a^{2}-s)^{j}\,\lambda^{1/2}(1,a^{2},s)\,\lambda^{1/2}(s,b^{2},c^{2})\nonumber \\
 &\left(\frac{1}{s}\lambda(s,b^{2},c^{2})+\frac{\left(1-a^{2}+s\right)^{2}}{2}\left(1+\displaystyle\frac{b^{2}+c^{2}}{s}-2\displaystyle\frac{(b^{2}-c^{2})^{2}}{s^{2}}\right)\right)\frac{\diff s}{s^{1+i/2}}\\
\nonumber \\
\Gamma_{3}^{ji}(a,b,c)&=4\int_{(b+c)^{2}}^{(1-a)^{2}} (1+a^{2}-s)^{j}\,\lambda^{1/2}(1,a^{2},s)\,\lambda^{1/2}(s,b^{2},c^{2})
\nonumber\\
&\left(\frac{a^{2}}{s}\lambda(s,b^{2},c^{2})+\frac{\left(1-a^{2}-s\right)^{2}}{2}\left(1+\displaystyle\frac{b^{2}+c^{2}}{s}-2\displaystyle\frac{(b^{2}-c^{2})^{2}}{s^{2}}\right)\right)\frac{\diff s}{s^{1+i/2}}\\
\nonumber \\
\Gamma_{4}^{ji}(a,b,c)&=4\int_{(b+c)^{2}}^{(1-a)^{2}} (1+a^{2}-s)^{j}\,\lambda^{1/2}(1,a^{2},s)\,\lambda^{1/2}(s,b^{2},c^{2})
\nonumber\\
&\left(\frac{(1+a^{2}-s)}{s}\lambda(s,b^{2},c^{2})+\left((1-a^{2})^{2}-s^{2}\right)\left(1+\displaystyle\frac{b^{2}+c^{2}}{s}-2\displaystyle\frac{(b^{2}-c^{2})^{2}}{s^{2}}\right)\right)\frac{\diff s}{s^{1+i/2}}
\end{align}
\endgroup

The full partial width for $N_h\to\nu_\ell\, l_a^-\,l_a^+$ is:
\begin{multline}
\Gamma(N_h\to\nu_\ell\,l^+_a l^-_a)=\Gamma_{\nu l^+l^-}^{W,\,{\rm SM}}
+\Gamma_{\nu l^+l^-}^{Z,\,{\rm SM}}
+\Gamma_{\nu l^+l^-}^{WZ,\,{\rm SM}}
+\Gamma_{\nu l^+l^-}^{Z,\,{\rm eff}}
+\Gamma_{\nu l^+l^-}^{\gamma,\,{\rm eff}}
+\Gamma_{\nu l^+l^-}^{Z\gamma,\,{\rm eff}} \\
+\Gamma_{\nu l^+l^-}^{WZ,\,{\rm SM+eff}}
+\Gamma_{\nu l^+l^-}^{W\gamma,\,{\rm SM+eff}}
+\Gamma_{\nu l^+l^-}^{Z,\,{\rm SM+eff}}
+\Gamma_{\nu l^+l^-}^{Z\gamma,\,{\rm SM+eff}}
\end{multline}
with:
\begingroup
\allowdisplaybreaks
\begin{eqnarray}
\Gamma_{\nu l^+l^-}^{W,\,{\rm SM}} &=&
\dfrac{G_F^{2}\,M_h^{5}}{96\pi^{3}}
\left|U_{a\ell}\,U_{ah}\right|^{2}\,\Gamma_{1}^{10}(x_a,x_a,0) \\
\Gamma_{\nu l^+l^-}^{Z,\,{\rm SM}} &=& \frac{G_F^{2}\,M_h^{5}}{96\pi^3}\left|C_{\ell h}\right|^{2}
\left(g_{L}^{l}\,g_{R}^{l}\,\Gamma_{5}^{10}(0,x_a,x_a)
+\left((g_{L}^{l})^{2}+(g_{R}^{l})^{2}\right)\Gamma_{1}^{10}(x_a,x_a,0)\right) \\
\Gamma_{\nu l^+l^-}^{WZ,\,{\rm SM}} &=&
\frac{G_F^{2}\,M_h^{5}}{96\pi^{3}}
\Re e \left[U_{a\ell}U_{ah}^{*}C_{\ell h}\right]
\left(2g_{L}^{l}\,\Gamma_{1}^{10}(x_a,x_a,0)
+g_{R}^{l}\Gamma_{5}^{10}(0,x_a,x_a)\right) \\
\Gamma_{\nu l^+l^-}^{Z,\,{\rm eff}} &=&
\frac{G_F\,s_{W}^{2}\,M_h^{7}}{24\sqrt{2}\pi^3 m_{Z}^{2}}\left|\frac{(\alpha'_{NB})_{\ell h}}{\Lambda}\right|^{2}
\Bigg(2g_{L}^{l}g_{R}^{l}\left(
3\Gamma_{5}^{10}(0,x_a,x_a)-\Gamma_{5}^{20}(0,x_a,x_a)\right) \nonumber \\
&&\begin{multlined}[b]
-\left((g_{L}^{l})^{2}+(g_{R}^{l})^{2}\right)\Big(\Gamma_{1}^{20}(0,x_{l_{1}},x_{l_{2}})+4\Gamma_{3}^{00}(0,x_a,x_a)\\
-\Gamma_{1}^{10}(0,x_a,x_a)-\Gamma_{2}^{10}(0,x_a,x_a)-\Gamma_{3}^{10}(0,x_a,x_a)-\Gamma_{4}^{10}(0,x_a,x_a)\Big)\Bigg)
\end{multlined} \nonumber \\ \\
\Gamma_{\nu l^+l^-}^{\gamma,\,{\rm eff}} &=&
\frac{\alpha_{\rm em}\,c_{W}^{2}\, M_h^{3}}{24\pi^2}\left|\frac{(\alpha'_{NB})_{\ell h}}{\Lambda}\right|^{2}
\Big(-\Gamma_{1}^{24}(0,x_a,x_a)-\Gamma_{5}^{24}(0,x_a,x_a) -4\Gamma_{3}^{04}(0,x_a,x_a) \nonumber \\
&&\begin{multlined}[b]
\hspace{8em}+\Gamma_{1}^{14}(0,x_a,x_a)+\Gamma_{2}^{14}(0,x_a,x_a)+\Gamma_{3}^{14}(0,x_a,x_a)\\
+\Gamma_{4}^{14}(0,x_a,x_a)+3\Gamma_{5}^{14}(0,x_a,x_a)\Big)
\end{multlined} \nonumber \\ \\
\Gamma_{\nu l^+l^-}^{Z\gamma,\,{\rm eff}} &=&
-\frac{\alpha_{\rm em}\,M_h^{5}}{24\pi^2 m_{Z}^{2}}\left|\frac{(\alpha'_{NB})_{\ell h}}{\Lambda}\right|^{2}(g_{L}^{l}+g_{R}^{l})
\Big(-\Gamma_{1}^{22}(0,x_a,x_a)-\Gamma_{5}^{22}(0,x_a,x_a)\nonumber \\
&&\begin{multlined}[b]
\quad-4\Gamma_{3}^{02}(0,x_a,x_a)+\Gamma_{1}^{12}(0,x_a,x_a)+\Gamma_{2}^{12}(0,x_a,x_a)+\Gamma_{3}^{12}(0,x_a,x_a)\\
+\Gamma_{4}^{12}(0,x_a,x_a)+3\Gamma_{5}^{12}(0,x_a,x_a)\Big)
\end{multlined} \nonumber \\
\Gamma_{\nu l^+l^-}^{WZ,\,{\rm SM+eff}} &=& 
-\frac{G_F\,e\,M_h^{6}}{96\sqrt{2}\pi^{3}c_{W}m_{Z}^{2}}
\Re e\left[\frac{(\alpha'_{NB})_{\ell h}}{\Lambda}U_{ah}U_{a\ell}^{*}\right]
\Big(3g_{R}^{l}\Gamma_{5}^{10}(0,x_a,x_a)\nonumber \\ 
&&\hspace{6em}+g_{L}^{l}\left(\Gamma_{1}^{10}(0,x_a,x_a)+\Gamma_{4}^{00}(0,x_a,x_a)-2\Gamma_{2}^{00}(0,x_a,x_a)\right)\Big) \nonumber \\ \\
\Gamma_{\nu l^+l^-}^{W\gamma,\,{\rm SM+eff}} &=&
\frac{G_F\,e\,c_{W}\,M_h^{4}}{96\sqrt{2}\pi^{3}}
\Re e\left[\frac{(\alpha'_{NB})_{\ell h}}{\Lambda}U_{ah}U_{a\ell}^{*}\right]
\Big(3\Gamma_{5}^{12}(0,x_a,x_a)+\Gamma_{1}^{12}(0,x_a,x_a) \nonumber\\
&&\hspace{15em}+\Gamma_{4}^{02}(0,x_a,x_a)-2\Gamma_{2}^{02}(0,x_a,x_a)\Big) \nonumber \\ \\
\Gamma_{\nu l^+l^-}^{Z,\,{\rm SM+eff}} &=&
-\frac{G_{F}\,e\,M_h^{6}}{96\sqrt{2}\pi^{3}c_{W}\,m_{Z}^{2}}
\Re e\left[\frac{(\alpha'_{NB})_{\ell h}}{\Lambda}C_{\ell h}\right]
\Big(6g_{L}^{l}\,g_{R}^{l}\,\Gamma_{5}^{10}(0,x_a,x_a)\nonumber \\
&&+\left((g_{L}^{l})^{2}+(g_{R}^{l})^{2}\right)\big(\Gamma_{4}^{00}(0,x_a,x_a)-2\Gamma_{3}^{00}(0,x_a,x_a)+\Gamma_{1}^{10}(0,x_a,x_a)\big)\Big) \nonumber \\ \\
\Gamma_{\nu l^+l^-}^{Z\gamma,\,{\rm SM+eff}} &=&
\frac{G_F\,e\,c_W\,M_h^{4}}{96\sqrt{2}\pi^3}
\Re e\left[\frac{(\alpha'_{NB})_{\ell h}}{\Lambda}C_{\ell h}\right]
(g_{L}^{l}+g_{R}^{l})\Big(\Gamma_{1}^{12}(0,x_a,x_a) \nonumber \\
&&\hspace{7em}-3\Gamma_{5}^{12}(0,x_a,x_a)+\Gamma_{4}^{02}(0,x_a,x_a)-2\Gamma_{3}^{02}(0,x_a,x_a)\Big)
\end{eqnarray}
We are defining $x_a\equiv m_{l_a}/M_h$.
\endgroup

The full partial width for $N_h\to\nu_\ell\, q\,\bar q$, after summing three quark colours, is:
\begin{equation}
\Gamma(N_h\to\nu\,q\,\bar q)=\Gamma_{\nu q\bar q}^{Z,\,\rm SM}+\Gamma_{\nu q\bar q}^{Z,\,\rm eff}+\Gamma_{\nu q\bar q}^{\gamma,\,\rm eff}+\Gamma_{\nu q\bar q}^{Z\gamma,\,{\rm eff}}
+\Gamma_{\nu q\bar q}^{Z,\,{\rm SM+eff}}+\Gamma_{\nu q\bar q}^{Z\gamma,\,{\rm SM+eff}}
\end{equation}
with:
\begin{align}
\Gamma_{\nu q\bar q}^{Z,\,\rm SM} &= 3\,(1+\Delta_{\rm QCD})\,\Gamma_{\nu l^+l^-}^{Z,\,{\rm SM}} &
\Gamma_{\nu q\bar q}^{Z,\,\rm eff} &= 3\,(1+\Delta_{\rm QCD})\,\Gamma_{\nu l^+l^-}^{Z,\,{\rm eff}} \\
\Gamma_{\nu q\bar q}^{\gamma,\,\rm eff} &= 
3\,(1+\Delta_{\rm QCD})\,Q_q^2\, \Gamma_{\nu l^+l^-}^{\gamma,\,{\rm eff}} &
\Gamma_{\nu q\bar q}^{Z\gamma,\,{\rm eff}} &= -3(1+\Delta_{\rm QCD})\,\,Q_q\,\Gamma_{\nu l^+l^-}^{Z\gamma,\,{\rm eff}} \\
\Gamma_{\nu q\bar q}^{Z,\,{\rm SM+eff}} &= 3\,(1+\Delta_{\rm QCD})\,\Gamma_{\nu l^+l^-}^{Z,\,{\rm SM+eff}}  &
\Gamma_{\nu q\bar q}^{Z\gamma,\,{\rm SM+eff}} &= -3\,(1+\Delta_{\rm QCD})\,Q_q\,\Gamma_{\nu l^+l^-}^{Z\gamma,\,{\rm SM+eff}} 
\end{align}
where in all equations above one should change $g_{L,R}^l\to g_{L,R}^q$ and $x_a\to x_q\equiv m_q/M_h$.

Finally, the full partial width for $N_h\to 3\nu_\ell$ is:
\begin{equation}
\Gamma(N_h\to3\nu_\ell)=
\Gamma_{3\nu_\ell}^{Z,\,\rm SM}
+\Gamma_{3\nu_\ell}^{Z,\,\rm eff}
+\Gamma_{3\nu_\ell}^{Z,\,{\rm SM+eff}}
\end{equation}
where we sum over all final light neutrinos. We have:
\begingroup
\allowdisplaybreaks
\begin{eqnarray}
\Gamma_{3\nu_\ell}^{Z,\,\rm SM}&=&
\frac{G_F^{2}\,M_h^{5}}{384\pi^{3}}\sum_{i,n,m}\left(\left|C_{ih}\,C_{nm}\right|^{2}
+\Re e\left[C_{ih}C_{nm}C_{nh}^{*}C_{im}^{*}\right]\right) \\
\Gamma_{3\nu_\ell}^{Z,\,\rm eff} &=&
\frac{G_F\,s^2_{W}\,M_{h}^{7}}{960\sqrt2\pi^{3}m_{Z}^{2}}
\sum_{i,n,m}\bigg(16\left|\frac{(\alpha'_{NB})_{ih}}{\Lambda}\,C_{nm}\right|^{2}
+6\left|\frac{(\alpha'_{NB})^*_{nm}}{\Lambda}\,C_{ih}\right|^{2} \nonumber \\
&&\begin{multlined}[b]
+4\,\Re e\left[\frac{(\alpha'_{NB})_{ih}}{\Lambda}\frac{(\alpha'_{NB})_{mh}^{*}}{\Lambda}C_{im}C_{nm}^{*}\right] 
-\Re e\left[\frac{(\alpha'_{NB})_{nm}^{*}}{\Lambda}\frac{(\alpha'_{NB})_{im}}{\Lambda}C_{ih}C_{nh}^{*}\right] \\
+12\,\Re e\left[\frac{(\alpha'_{NB})_{ih}}{\Lambda}\frac{(\alpha'_{NB})_{im}^{*}}{\Lambda}C_{nh}^*C_{nm}\right]\bigg)
\end{multlined} \nonumber \\ \\
\Gamma_{3\nu_\ell}^{Z,\,{\rm SM+eff}} &=&
\frac{G_F\,e\,M_{h}^{6}}{192\sqrt2\pi^{3}c_W\,m_{Z}^{2}}
\sum_{i,n,m}\bigg(
\left|C_{nm}\right|^{2}\Re e\left[\frac{(\alpha'_{NB})_{ih}}{\Lambda}C_{ih}\right]\nonumber \\
&&\hspace{16em}
+\Re e\left[\frac{(\alpha'_{NB})_{nh}}{\Lambda}C_{ih}C_{nm}C_{im}^{*}\right]\bigg) \nonumber \\
\end{eqnarray}
\endgroup 

\section{Implementation of cuts in LEP analysis}
\label{app:LEP}

After analytically calculating the differential $e^+e^-\to N_h\,\nu_\ell$ cross section, $d\sigma_{N\nu}/d\Omega$, the total cross section is obtained by integrating over the solid angle. It is at this point that the heavy neutrino lifetime constraint, as well as photon energy and angle cuts are applied.

Our final formula is:
\begin{multline}
\sigma_{N\nu}^{\rm cuts}=\frac{(\hbar c)^2}{32\pi \,m_Z^2}\left(1-\frac{M_h^2}{m_Z^2}\right)
\left(\frac{1}{4\pi\,\tau_{N}^{\rm lab}}\right){\rm BR}(N_h\to\nu\,\gamma) \\
\int d(\cos\theta_{\gamma})\,d\phi_\gamma\,d(\cos\theta_N)\,dt_N\exp\left[-\frac{t_N}{\tau_{N}^{\rm lab}}\right]\frac{d\sigma_{N\nu}}{d\cos\theta_N}\Theta_H\left(\sqrt{x_\gamma^2+y_\gamma^2}-z_{\rm det}\tan\theta_{\rm veto}\right)
\end{multline}
where $\theta_\gamma$ and $\phi_\gamma$ are the $\vec p_\gamma$ angles in the heavy neutrino rest frame, with respect to the direction of the heavy neutrino momentum, and $\theta_N$ is the $\vec p_N$ angle with respect to the beamline. Since the heavy neutrinos are Majorana particles, we assume the photon distribution to be isotropic in the $N_h$ rest frame. The photon energy cut $E_\gamma^{\rm cut}=0.7\,$GeV is satisfied if:
\begin{eqnarray}
\cos\theta_\gamma>\frac{1}{\beta_{\rm rel}}\left(\frac{2E_\gamma^{\rm cut}}{\gamma_{\rm rel} M_h}-1\right)
\end{eqnarray}
where $\beta_{\rm rel}$ and $\gamma_{\rm rel}$ are the relativistic boost parameters. The heavy neutrino time of flight $t_N$ is integrated up to $t_N^{\rm max}=d_N^{\rm max} E_N/(|\vec p_N|c)$, with $d_N^{\rm max}=2\,$m. The heavy neutrino lifetime in the lab frame is defined as $\tau_N^{\rm lab}\equiv(E_N/M_h)\times(\hbar/\Gamma_N)$. Finally, the photon is required to enter the electromagnetic calorimeter of the experiment, which is assumed to be the HPC of DELPHI~\cite{Chan:1995urq}. This was taken in~\cite{Magill:2018jla,Lopez:1996ey} as a cut on the photon angle with respect to the beamline, $\cos\theta_{\rm veto}>0.7$. In order to apply this bound, for a given $t_N$ and $\theta_N$ we determine the position where the heavy neutrino decayed, $d_N$, and trace the photon trajectory by boosting its momentum in the heavy neutrino direction. The photon thus moves along the line:
\begin{equation}
\vec r_\gamma=\begin{pmatrix}
d_N\,s_{\theta_N} \\ 0 \\ d_N\,c_{\theta_N}
\end{pmatrix}+R
\begin{pmatrix}
s_{\theta_N}c_{\theta'_\gamma}+c_{\theta_N}s_{\theta'_\gamma}c_{\phi_\gamma} \\
s_{\theta'_\gamma}s_{\phi_\gamma} \\
c_{\theta_N}c_{\theta'_\gamma}-s_{\theta_N}s_{\theta'_\gamma}c_{\phi_\gamma}
\end{pmatrix}
\end{equation}
with $\theta'_\gamma$ equal to the $\vec p_\gamma$ angle with respect to the heavy neutrino momentum in the lab frame:
\begin{eqnarray}
\cos\theta'_\gamma=\frac{\gamma_{\rm rel}(\cos\theta_\gamma+\beta_{\rm rel})}{\sqrt{(\gamma_{\rm rel}\cos\theta_\gamma+\beta_{\rm rel}\gamma_{\rm rel})^2+\sin^2\theta_\gamma}}
\end{eqnarray}

Unless it is absorbed by the calorimeter, the photon would reach the edge of the HPC at $z_{\rm det}=248\,$cm for
\begin{equation}
R_{\rm edge}=\frac{\pm z_{\rm det}-d_N c_{\theta_N}}{c_{\theta_N}c_{\theta'_\gamma}-s_{\theta_N}s_{\theta'_\gamma}c_{\phi_\gamma}}\geq0
\end{equation}
Thus, if we now define $x_\gamma=d_N s_{\theta_N}+R_{\rm edge}(s_{\theta_N}c_{\theta'_\gamma}+c_{\theta_N}s_{\theta'_\gamma}c_{\phi_\gamma})$ and $y_\gamma=R_{\rm edge}\,s_{\theta'_\gamma}s_{\phi_\gamma}$, requiring the photon to leave the detector through the calorimeter amounts to demanding $\sqrt{x_\gamma^2+y_\gamma^2}>z_{\rm det}\tan\theta_{\rm veto}$, which is implemented in the integral above via a Heaviside $\Theta_H$ function.

\bibliographystyle{JHEP}
\bibliography{main}

\end{document}